\documentclass[useAMS,usenatbib]{mn2e}
\usepackage{graphicx}
\usepackage{amsmath}
\usepackage{multirow}
\usepackage{rotating} 
\def\fluxunits{erg\,s$^{-1}$\,cm$^{-2}$}
\def\lumin{erg\,s$^{-1}$}
\title[SAX J1818.6$-$1703 Spectral Variation]{Spectral variation in the supergiant fast X-ray transient SAX J1818.6$-$1703 observed by \emph{XMM--Newton} and \emph{INTEGRAL}}
\author[C. M. Boon et al.]
	{C. M. Boon$^1$\thanks{C.M.Boon@soton.ac.uk}, A. J. Bird$^{1}$, A. B. Hill$^{1}$, L. Sidoli$^{2}$, V. Sguera$^{3}$, M. E. Goossens$^{1}$, \and  M. Fiocchi$^{4}$, V. A. McBride$^{5,6}$, S. P. Drave$^{1}$
        \\
	$^1$School of Physics and Astronomy, University of Southampton, University Road, Southampton, SO17 1BJ, UK \\
	$^2$INAF-IASF, Istituto di Astrofisica Spaziale e Fisica Cosmica, Via E. Bassini 15, I-20133 Milano, Italy \\
	$^3$INAF-IASF, Istituto di Astrofisica Spaziale e Fisica Cosmica, Via Gobetti 101, I-40129 Bologna, Italy \\
	$^4$INAF, Istituto di Astrofisica e Planetologia Spaziali, Via Fosso del Cavaliere 100, I-00133 Roma, Italy \\
	$^5$Department of Astronomy, Astrophysics, Cosmology and Gravity Centre, University of Cape Town, \\
	 \, Private Bag X3, Rondebosch 7701, South Africa \\
	$^6$South African Astronomical Observatory, PO Box 9, Observatory 7935, South Africa 
}
	
\begin{document}

\date{Accepted 2015 December 19;  Received 2015 December 9; in original form 2015 February 9}

\pagerange{\pageref{firstpage}--\pageref{lastpage}} \pubyear{2014}

\maketitle

\label{firstpage}

\begin{abstract}
We present the results of a 30\,ks \emph{XMM--Newton} observation of the supergiant fast X-ray transient (SFXT) SAX J1818.6$-$1703 $-$ the first in-depth soft X-ray study of this source around periastron. \emph{INTEGRAL} observations shortly before and after the \emph{XMM--Newton} observation show the source to be in an atypically active state. Over the course of the  \emph{XMM--Newton} observation, the source shows a dynamic range of $\sim 100$ with a luminosity greater than $1\times10^{35}$\, \lumin ~for the majority of the observation. After an $\sim$6\,ks period of low luminosity ($\sim10^{34}$\, \lumin) emission, SAX J1818.6$-$1703 enters a phase of fast flaring activity, with flares $\sim$250\,s long, separated by $\sim$2\,ks. The source then enters a larger flare event of higher luminosity and $\sim$8\,ks duration. Spectral analysis revealed evidence for a significant change in spectral shape during the observation with a photon index varying from $\Gamma \, \sim \, 2.5$ during the initial low luminosity emission phase,  to $\Gamma \, \sim \, 1.9$  through the fast flaring activity, and a significant change to $\Gamma \, \sim \, 0.3$ during the main flare. The intrinsic absorbing column density throughout the observation (n$_{H} \, \sim \, 5 \times 10^{23}$ cm$^{-2}$) is among the highest measured from an SFXT, and together with the \emph{XMM--Newton} and \emph{INTEGRAL} luminosities, consistent with the neutron star encountering an unusually dense wind environment around periastron. Although other mechanisms cannot be ruled out, we note that the onset of the brighter flares occurs at $3\times10^{35}$\lumin, a luminosity consistent with the threshold for the switch from a radiative-dominated to Compton cooling regime in the quasi-spherical settling accretion model. 

\end{abstract}


%%%%%%%%%%%%%%%%%%%%%%%%%%%%%%%%%%%%%%%%%%%%%%%%

% Introduction

%%%%%%%%%%%%%%%%%%%%%%%%%%%%%%%%%%%%%%%%%%%%%%%%

\section{Introduction}
Supergiant fast X-ray transients (SFXTs) are a subclass of high mass X-ray binaries (HMXBs) recognised over the course of the \emph{INTEGRAL} mission (\citealt{Sguera2005}, \citealt{2006ApJ...646..452S}, \citealt{2006ESASP.604..165N}). These sources display a large X-ray dynamic range, with intermediate SFXTs such as IGR J16465$-$4514 showing a dynamic range  of 100 \citep{Clarke2010} while extreme examples, such as IGR J17544$-$2619, possess an observed dynamic range of $10^{6}$ from quiescence \citep{2005A&A...441L...1I} to maximum flare luminosity \citep{2015A&A...576L...4R}. There are currently 14 spectroscopically confirmed SFXTs lying along the Galactic plane and approximately the same number of candidate systems exist, though they lack spectroscopic confirmation of a supergiant companion star. 

A number of geometric effects have been proposed in order to explain the rapid transient nature of SFXTs, including: clumping in the spherically symmetric outflow of the supergiant companion \citep{2005A&A...441L...1I}, an enhancement of the equatorial density of the supergiant wind inclined with respect to the orbit of the neutron star \citep{2007A&A...476.1307S} and variations in orbital period and eccentricity to distinguish between transient and persistent supergiant systems \citep{2008AIPC.1010..252N}. However, \citet{2007AstL...33..149G} have argued that observed outburst profiles could not be explained solely by the presence of an individual density inhomogeneity.

More complicated models invoked to explain such a large dynamic range in these systems make use of transitions in accretion mechanisms. \citet{2008ApJ...683.1031B} suggested transitions between gating mechanisms, such as centrifugal and magnetic barriers, initiated by density variations in the stellar wind could cause the observed flux ranges.

Recently, \cite{Shakura2012a} have proposed a model of quasi-spherical accretion on to slowly rotating neutron stars, in which accretion on to the neutron star is mediated through a quasi-static shell of plasma above the magnetosphere. In this model, bright flares are caused by a transition from a less efficient radiative cooling mechanism to a more efficient Compton cooling regime. The transition to a more effective cooling regime allows the plasma reservoir to drain on to the neutron star through the magnetosphere precipitating a rapid increase in luminosity. Originally applied to the slowly rotating pulsars Vela X$-$1, GX 301$-$2 and 4U 1907$+$09 \citep{Shakura2012a}, it has since been used to explain the fast flaring behaviour seen in SFXTs (\citealt{2013MNRAS.433..528D}; \citealt{2014MNRAS.439.2175D}; \citealt{Paizis2014}; \citealt{2014MNRAS.442.2325S}).

SAX J1818.6$-$1703 was discovered by the \emph{BeppoSAX} observatory on 1998 March 11 in a short X-ray outburst lasting a few hours \citep{1998IAUC.6840....2I}. Renewed short X-ray activity was subsequently detected by the IBIS/ISGRI telescope on \emph{INTEGRAL} on 2003 September 9 as two intense short outbursts reaching $\sim$380 mCrab in the 18--45 keV energy band \citep{2005AstL...31..672G}. The observed activity of SAX J1818.6$-$1703 by \citet{2005AstL...31..672G} showed complex time varying behaviour lasting approximately 1\,d, with the main flare event lasting 2.7\,h. IBIS/ISGRI spectra taken from this observation were fitted well with a thermal bremsstrahlung model, showing signs of spectral evolution through the flaring event. The nature of the system was unknown at the time of this observation, though the authors did rule out changes in accretion through a standard accretion disc as the cause of the behaviour. Other short X-ray outbursts from SAX J1818.6$-$1703 have been also observed with \emph{RXTE} \citep{2012MNRAS.422.2661S}.

The identification of a B0.5Iab supergiant within the error circle of the refined \emph{Chandra} position of SAX J1818.6$-$1703 identified the system as an SFXT at a distance of 2.1\,kpc  (\citealt{2006ATel..831....1N}; \citealt{2010A&A...510A..61T}).

The source was undetected in a 13 ks observation on 2006 October 8 with \emph{XMM--Newton} by \citet{2008ATel.1493....1B}. The authors placed a 3$\sigma$ upper limit on the unabsorbed 0.5--10 keV X-ray flux of 1.1 $\times$ 10$^{-13}$ \fluxunits (L$_{X} \, = \, 5.8 \, \times \, 10^{31}$ ergs\,s$^{-1}$ at a distance of 2.1\,kpc) using data from the EPIC-pn camera. The source was also not detected in a 45 ks observation on 2010 March 21 with \emph{XMM} \citep{2012A&A...544A.118B}, though due to high flaring background, the effective observation time for EPIC-pn was only 4 ks. This observation placed a 3$\sigma$ upper limit on the unabsorbed 0.5--10 keV flux of 3 $\times$ 10$^{-13}$ \fluxunits, consistent with the previous observations. 

An in depth study of available \emph{INTEGRAL}/IBIS data by \citet{Bird2008} revealed a 30$\pm$0.1\,d orbital period. A period of 30$\pm$0.2\,d was independently found by \citet{Heras2008} and the eccentricity of the orbit was constrained to between 0.3 and 0.4. Using both the orbital period and the ephemeris (MJD 54540.659) from \citet{Bird2008}, the previous \emph{XMM} observations were found to have been taken at orbital phases of $\phi = 0.51$ and $0.53$ (close to apastron) for the 13 and 45 ks observations respectively. Using the same \emph{INTEGRAL} data set and performing a recurrence analysis of outbursts from SAX J1818.6$-$1703, \citet{Bird2008} found that the system has a high level of recurrence, showing detectable emission in more than 50 \% of periastron passages. This high recurrence rate when coupled with the large outbursts as seen by \citet{2005AstL...31..672G} show the behaviour of SAX J1818.6$-$1703 is atypical of that found in other SFXTs.

A catalogue of {\it Swift}/BAT bright flares from the SAX J1818.6$-$1703 has been reported by \citet{2014A&A...562A...2R}, while a systematic re-analysis of the \emph{INTEGRAL} archival data has been reported by \citet{Paizis2014}, who found that the distribution of X-ray flare luminosities follow a power-law.

We present phase targeted temporal and spectral analysis of this source, observed with \emph{XMM--Newton} starting on 2013 March 21 and lasting 30 ks. Spectral analysis of this source shows evidence for variation in both absorbing column density and photon index correlated with flaring behaviour. 

%%%%%%%%%%%%%%%%%%%%%%%%%%%%%%%%%%%%%%%%%%%%%%%%

% Data Analysis

%%%%%%%%%%%%%%%%%%%%%%%%%%%%%%%%%%%%%%%%%%%%%%%%
\section{Data Analysis}
\label{sec:Data}

%%%%%%%%%% Table: Spectrum Timestamps
\begin{table*}
\begin{center}
\caption{Log of observations of SAX J1818.6$-$1703 discussed in this work. Phases are calculated using the ephemeris of \citet{Bird2008}. }
\begin{tabular}{cllccc}
\hline
\hline
Instrument 			& UTC  						& MJD 			& Phase 			& Exposure (ks) & Satellite Revolution \\
\hline
\emph{INTEGRAL}/IBIS& 2013 March 12 00:19:37 --  & 56363.014 -- & 0.804\,--\,0.862  & 10 			& 1271 \\	
 					& 2013 March 13 18:15:52 	 & 56364.761    &					& 				&      \\[4pt]
\emph{INTEGRAL}/IBIS& 2013 March 20 11:10:03 --  & 56371.465 -- & 0.086\,--\,0.099  & 7.5			& 1274 \\
					& 2013 March 20 20:44:48	 & 56371.864    & 					& 				& 	   \\[4pt]	
\emph{XMM--Newton}	& 2013 March 21 12:37:22 --  & 56372.514 -- & 0.121\,--\,0.134  & 30  			& 2432 \\
					& 2013 March 21 21:24:54	 & 56372.914 	& 					& 				&      \\[4pt]
\emph{INTEGRAL}/IBIS& 2013 March 22 09:00:11 --  & 56373.375 -- & 0.150\,--\,0.165  & 7.5 			& 1274 \\
					& 2013 March 22 20:13:27 	 & 56373.843 	& 					& 				&      \\
\hline
\hline
\end{tabular}
\label{tab:ObsLog}
\end{center}
\end{table*}
%%%%%%%%%% End: Table

\subsection{\emph{XMM} data analysis}
\label{sec:XMM_Data}

%%%%%%%%%% Figure: IBIS maps
\begin{figure*}
\begin{center}
\includegraphics[width=0.47\textwidth,natwidth=583,natheight=580]{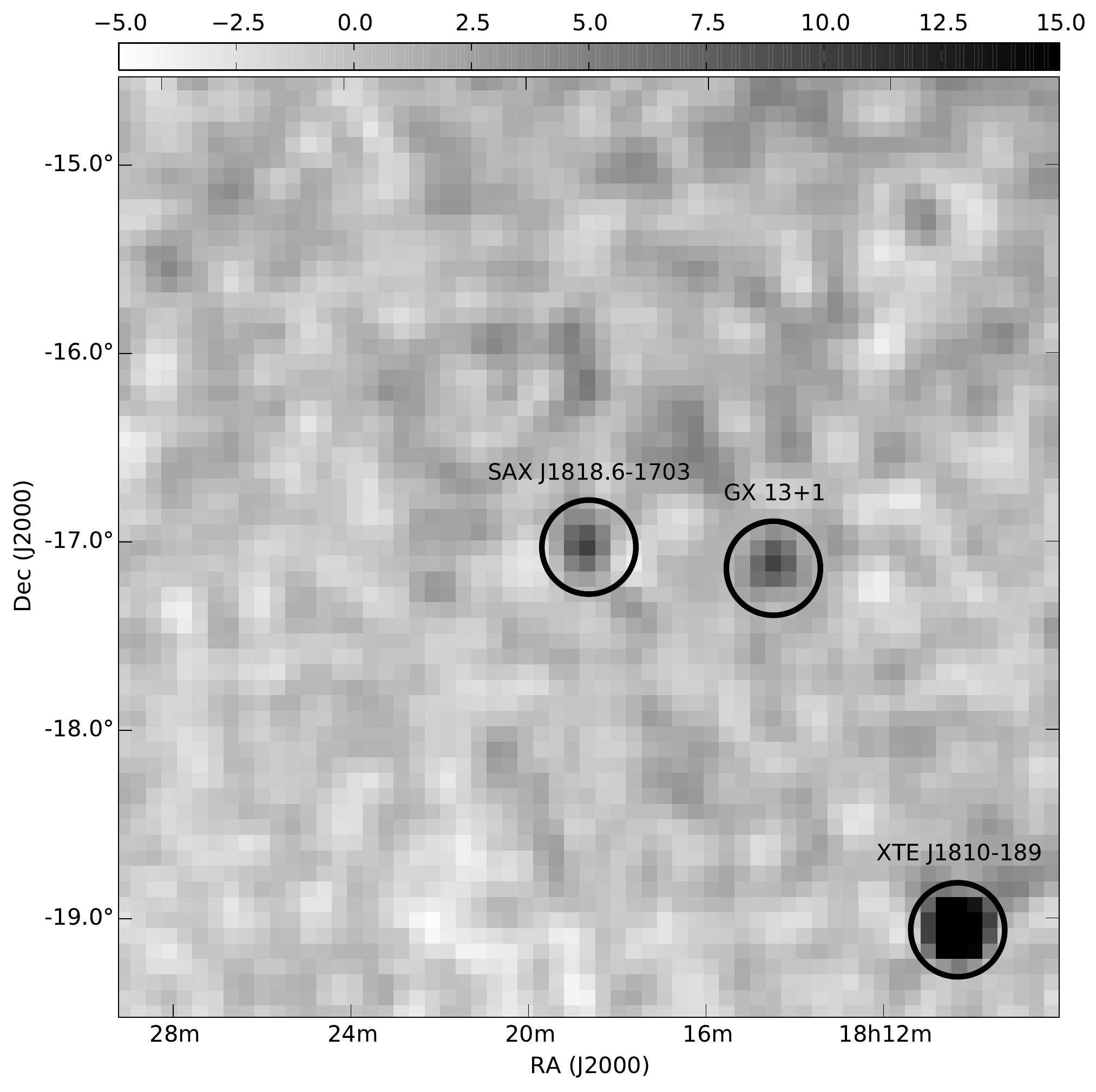}
\includegraphics[width=0.47\textwidth,natwidth=583,natheight=580]{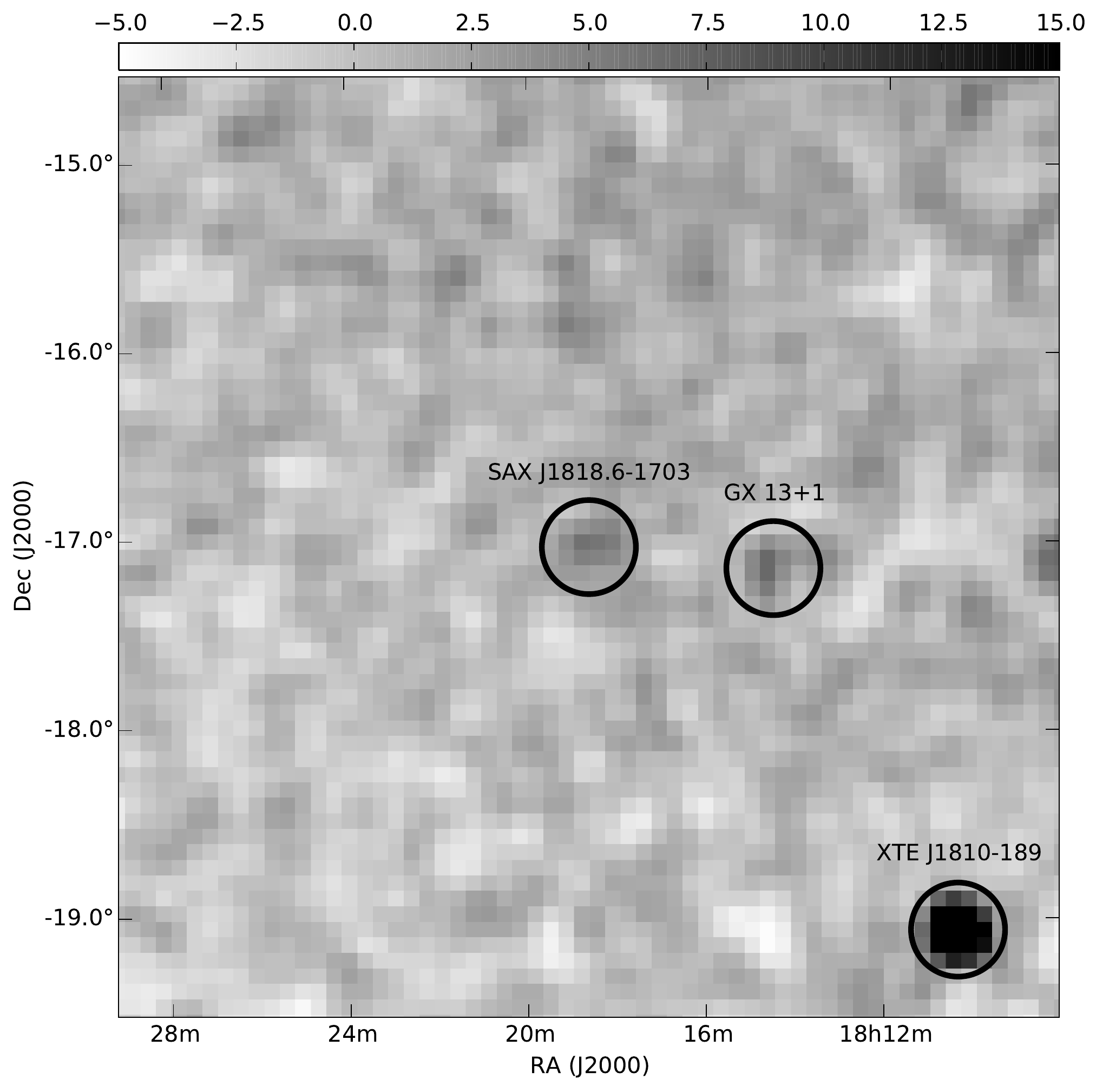}
\caption{ Left-hand panel: 18--60\,keV \emph{INTEGRAL}/IBIS significance map from the observation of SAX J1818.6$-$1703 between  UTC 11:10:03 and UTC 20:44:48 on 2013 March 20. The best X-ray positions of SAX J1818.6$-$1703 and other sources in the field of view are marked with black circles.  Right-hand panel: as the left-hand panel for the observation between UTC 09:00:11 and UTC 20:13:27 on 2013 March 22.  }
\label{fig:INT_maps}
\end{center}
\end{figure*}
%%%%%%%%%% End: Figure

A single \emph{XMM--Newton}/EPIC (\citealt{2001A&A...365L...1J}; \citealt{2001A&A...365L..27T}; \citealt{2001A&A...365L..18S}) observation of SAX J1818.6$-$1703 around the periastron orbital phase of the system was performed beginning at UTC 12:37:22 and ending at UTC 21:24:54 on 2013 March 21, covering the orbital phase region $\phi$\,$=$\,0.121\,--\,0.134 using the ephemeris of \citet{Bird2008} (ObsID: 0693900101). The EPIC-pn camera was operated in small window mode, while the EPIC-MOS cameras were operated in full frame mode. This configuration was chosen to account for the rapid transient nature of this source and reduce pile up in the event of the source experiencing a bright ($\sim$10$^{-10}$ \fluxunits) outburst.

Data from both EPIC-pn and EPIC-MOS detectors were analysed using the standard \emph{XMM--Newton} Science Analysis System (SAS v12.0.1; \citealt{2004ASPC..314..759G}) and the most recent instrument calibration files. The standard {\scshape epproc} and {\scshape emproc} tasks were used to produce cleaned event files for EPIC-pn and EPIC-MOS cameras, respectively. The data sets were filtered for flaring particle background and checked for photon pileup using the {\scshape epatplot} tool as described in the \emph{XMM--Newton} SAS data analysis threads\footnote{http://xmm.esac.esa.int/sas/current/documentation/threads/}. The observation was found not to suffer from photon pile up. We created single-event products ({\scshape pattern = 0}) in the range of 10--12 keV for EPIC-pn and above 10 keV for EPIC-MOS cameras covering the full field of view for each camera and using cut offs of 0.35 and 0.4 counts s$^{-1}$ for EPIC-pn and EPIC-MOS respectively as outlined in the SAS threads. Filtered event files were produced from the resulting good time intervals defined by this filtering. 

Optimal extraction regions for all light curve and spectrum generating procedures were generated using the SAS tool {\scshape eregionanalyse}, which yielded extraction regions of 26, 51 and 58 arcsec for EPIC-pn, EPIC-MOS1 and EPIC-MOS2 respectively.

Broad-band 0.2-10 keV light curves from EPIC-pn and EPIC-MOS were extracted with 100s resolution initially, as shown in Fig. \ref{fig:PNlc}. The properties of the EPIC-pn light curve are discussed in Section \ref{sec:XMMtemporal}. 

All spectra reported below were extracted using standard procedures from the \emph{XMM--Newton} analysis threads and the appropriate response for each spectrum was generated using the SAS tools {\scshape rmfgen} and {\scshape arfgen}. The spectra were fit in the 0.5-15\,keV band using {\scshape xspec} version 12.7.1 \citep{1996ASPC..101...17A} and uncertainties are quoted at the 90\% confidence level. In this work, the interstellar element abundances for photoelectric absorbing model components are set by the abundances of \citet{2000ApJ...542..914W}. Spectral properties are discussed in Section \ref{sec:spectral}.

\subsection{\emph{INTEGRAL} data analysis}
\label{sec:INTEGRAL_Data}

\emph{INTEGRAL}/IBIS (\citealt{2003A&A...411L.131U}; \citealt{2003A&A...411L...1W}) observed SAX J1818.6$-$1703 between UTC 00:19:37 on 2013 March 12 and UTC 18:15:52 2013 March 13 as part of Revolution 1271. During this revolution, the source was observed for a total of 10\,ks. SAX J1818.6$-$1703 was also observed by \emph{INTEGRAL}/IBIS between UTC 11:10:03 and UTC 20:44:48 on 2013 March 20 and between UTC 09:00:11 and UTC 20:13:27 on 2013 March 22. These \emph{INTEGRAL} observations cover the phases $\phi$\,$=$\,0.804\,--\,0.862, $\phi$\,$=$\,0.086\,--\,0.099 and $\phi$\,$=$\,0.150\,--\,0.165 respectively, using the ephemeris of \cite{Bird2008}. This archival data were processed and analysed using the \emph{INTEGRAL} Offline Science Analysis \citep{2003A&A...411L.223G} software version 10.1. Images were created in the 18--60\,keV energy range for each science window (ScW; an individual pointing of \emph{INTEGRAL} lasting approximately 2\,ks) and each observation of the source. Light curves in the 18--60\,keV band on ScW and 250\,s time-scales and spectra in the 18--500\,keV band were also generated following standard procedures. A list of all observations discussed in this work can be found in Table \ref{tab:ObsLog}. We are aware that the low energy threshold of \emph{IBIS/ISGRI} on-board \emph{INTEGRAL} has increased since launch and consequently the \emph{INTEGRAL} Science Data Centre recommend ignoring data below 22\,keV for all revolutions after Revolution 1090. We generated images in the 18--60\,keV and 22--60\,keV bands for both observations in Revolution 1274. We find that there is no significant difference in the detections of the source between the energy bands. In order to compare our results with previous hard X-ray studies using \emph{INTEGRAL}, we adopt the 18--60\,keV energy band for subsequent analysis.

%%%%%%%%%%%%%%%%%%%%%%%%%%%%%%%%%%%%%%%%%%%%%%%%

% Results

%%%%%%%%%%%%%%%%%%%%%%%%%%%%%%%%%%%%%%%%%%%%%%%%
\section{Results}
\label{sec:Results}
%%%%%%%%%%%%%%%%%%%%%%%%%%%%%%%%%%%%%%%%%%%%%%%%

% INTEGRAL context

%%%%%%%%%%%%%%%%%%%%%%%%%%%%%%%%%%%%%%%%%%%%%%%%

\subsection{\emph{INTEGRAL} results}
\label{sec:INTEGRAL_context}

SAX J1818.6$-$1703 was not detected in the earlier observation of the source during Revolution 1271, which was prior to the periastron passage. Fig. \ref{fig:INT_maps} shows the \emph{INTEGRAL}/IBIS 18--60\,keV significance maps for the observations in Revolution 1274 that bracket the \emph{XMM} observation of SAX J1818.6$-$1703. The left-hand panel shows the significance map for the observation taken one day prior to the \emph{XMM} observation; SAX J1818.6$-$1703 is detected at the 10$\sigma$ level. The right-hand panel shows the significance map for the observation taken one day after the \emph{XMM} observation. In this map, there is a marginal detection of the source at 6$\sigma$, however due to the short exposure time, there is a high level of systematic noise in the map. To look for signatures of fast variability the 250\,s bin light curve of these two observations was analysed. There is evidence of activity and variability through this observation period.

%%%%%%%%%%%%%%%%%%%%%%%%%%%%%%%%%%%%%%%%%%%%%%%%

% XMM Temporal	

%%%%%%%%%%%%%%%%%%%%%%%%%%%%%%%%%%%%%%%%%%%%%%%%
\subsection{XMM Temporal Analysis}
\label{sec:XMMtemporal}
\begin{figure}
\begin{center}
\includegraphics[width=0.5\textwidth,natwidth=576,natheight=432]{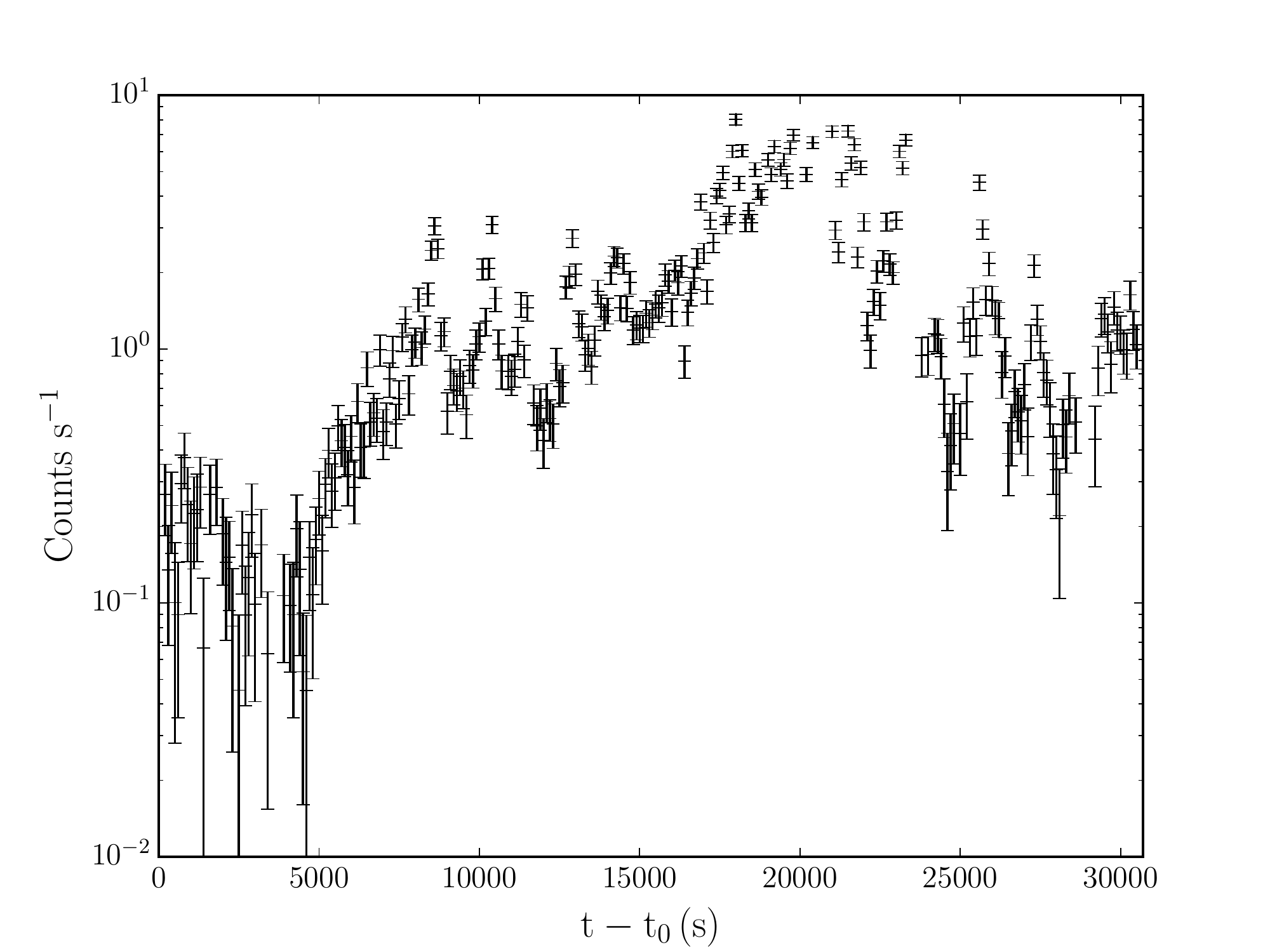}
\caption{0.2-10 keV EPIC-pn 100\,s resolution light curve of the \emph{XMM} observation of SAX J1818.6$-$1703. In this case, t$_{0}\,=\,\mathrm{MJD} \,56372.530$  and is the first time stamp of the EPIC-pn light curve. This observation covered the orbital phase region $\phi$\,$=$\,0.121\,--\,0.134.}
\label{fig:PNlc}
\end{center}
\end{figure}

Fig. \ref{fig:PNlc} shows the background subtracted 0.2--10 keV EPIC-pn light curve with 100\,s time bins and filtered for low fractional exposure ({\scshape fracexp} $<$ 0.5). Times in this light curve are given with  t$_{0}\,=\,\mathrm{MJD} \,56372.530$ being the first time stamp of the EPIC-pn light curve.  For t $<$  6000\,s, the source shows a low level of X-ray emission with an average flux for this region of the light curve is 0.16$\pm$0.009 counts s$^{-1}$ and a minimum count rate of 0.01$\pm$0.04 counts s$^{-1}$. Following this, the source then shows an increase in activity up until t$\sim$7500\,s when the source enters a period lasting $\sim$7.5\,ks where 3 small flares occur, reaching up to counts of 2-3 counts s$^{-1}$. Following the onset of this flaring activity, a large (up to $\sim$8 counts s$^{-1}$) flare occurs, lasting approximately 7.5\,ks after which the source exhibits small-scale flaring activity until the end of the observation. Over the course of the observation, the source exhibits a dynamic range of $\sim$100 in count rate.

%%%%%%%%%% Figure: PN light curve
%%%%%%%%%% End: Fig

%%%%%%%%%% Figure: HR vs T
\begin{figure}
\begin{center}
\includegraphics[width=0.5\textwidth,natwidth=683,natheight=370]{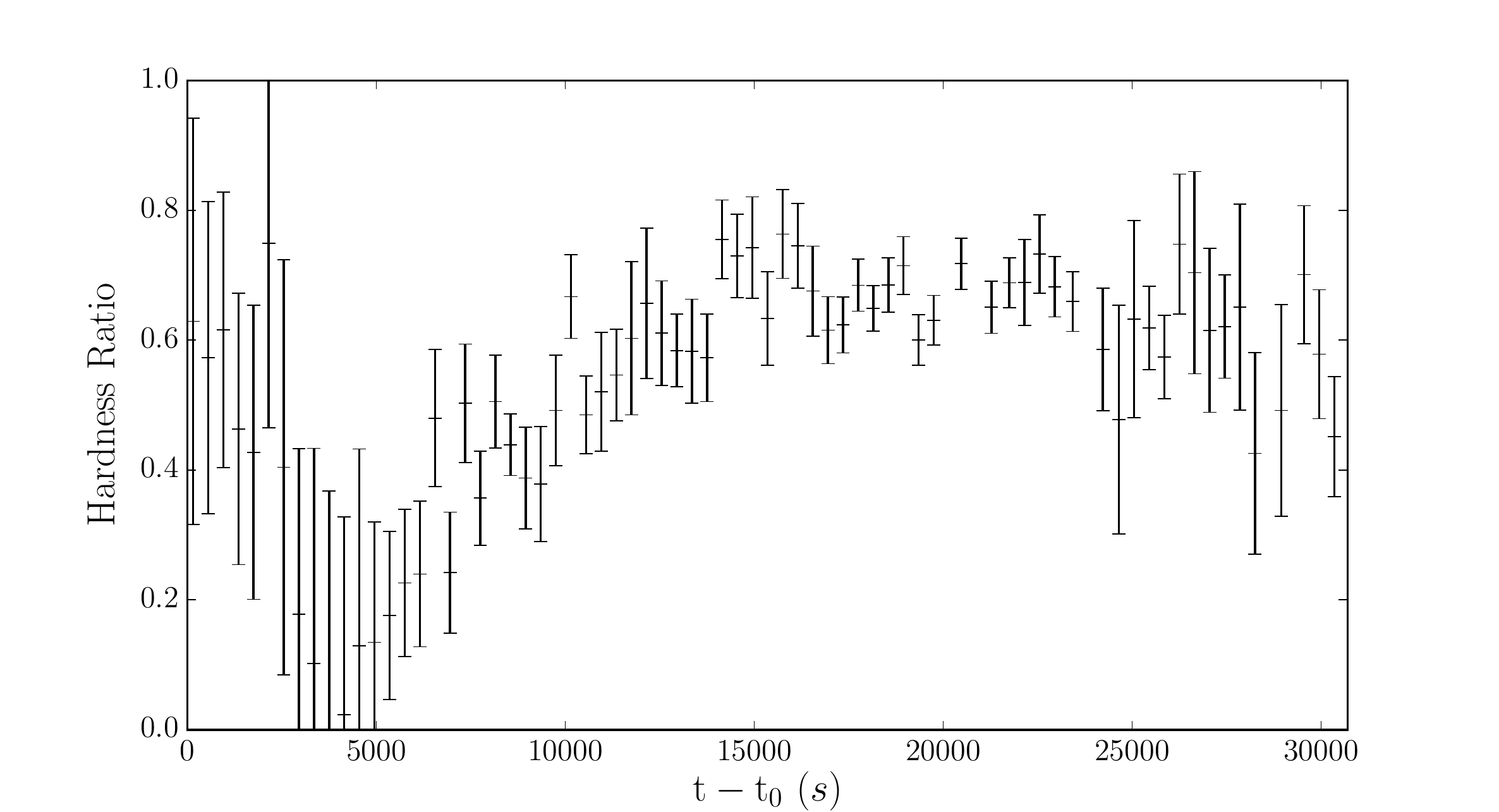}
\caption{ Hardness ratio of the 0.5--5\,keV and 5--10\,keV EPIC-pn light curves of SAX J1818.6$-$1703 with 400\,s binning. As with Fig. \ref{fig:PNlc}, t$_{0}\,=\,\mathrm{MJD} \,56372.530$  and is the first time stamp of the EPIC-pn light curve.}
\label{fig:HRvsT}
\end{center}
\end{figure} 
%%%%%%%%%% End: Figure

%%%%%%%%%% Figure: PN Light curve with hardness ratio
\begin{figure}
\begin{center}
\includegraphics[width=0.5\textwidth,natwidth=576,natheight=432]{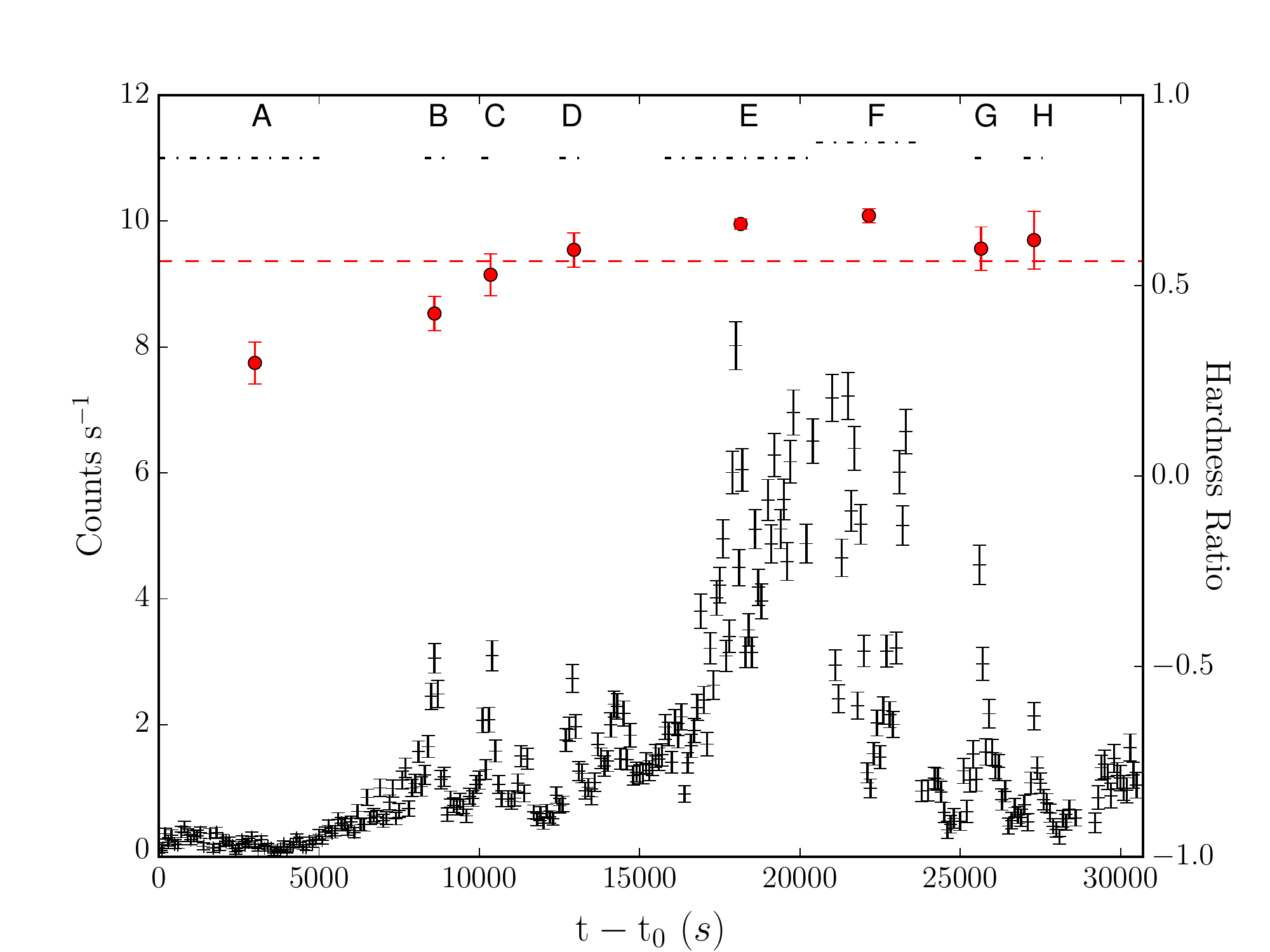}
\caption{ EPIC-pn 0.2--10\,keV 100\,s resolution light curve of SAX J1818.6$-$1703 as shown in Fig. \ref{fig:PNlc}. Regions used for spectral extractions are labelled A--H and the extent of these regions shown by the dot-dashed lines The hardness ratio for each of the regions labelled at the top of the figure are shown as red circles. The dashed line corresponds to the hardness ratio for the time intervals not covered by the labelled regions.}
\label{fig:LCwHR}
\end{center}
\end{figure} 
%%%%%%%%%% End: Figure

The EPIC-MOS1 and EPIC-MOS2 light curves show similar levels of variability and flaring activity, though due to the lower sensitivity of these instruments and hence lower count rates with respect to the EPIC-pn instrument, we focus our temporal analysis on the EPIC-pn light curves for the remainder of this work. 

A filtered light curve with 1\,s binning was tested for the presence of periodicities in an attempt to locate the as yet unknown neutron star spin period of SAX J1818.6$-$1703. A Lomb--Scargle periodogram (\citealt{1976Ap&SS..39..447L}; \citealt{1982ApJ...263..835S}) was produced using the fast implementation of \citet{1989ApJ...338..277P}. Significance levels, above which one would consider the detection of a periodic signal to be true, were calculated using Monte Carlo simulations. The method of \cite{2005A&A...439..255H} was used in order to gain appropriate confidence levels with 100 000 iterations utilised. Using this method when applied to the light curve of the whole observation, we did not find any significant peaks in the Lomb--Scargle power in the range 2\,s -- 4.2\,h. Searches for periodic signals in temporally selected windows corresponding to low and brighter intensities (t $<$ 8500, 8500 $<$ t $<$ 15\,000, 15\,000 $<$ t $<$ 25\,000, t $>$ 25\,000 ) were also carried out in the range of 2--2500\,s and yielded no significant signals. 

An epoch folding method using the Q$^{2}$ statistic as defined by \cite{1983ApJ...266..160L} was also utilised to identify periodic signals.
As for the Lomb-Scargle periodogram, 100,000 iterations of a Monte Carlo process were undertaken and confidence levels determined. This method yielded no significant peaks in the range 2s--4.2\,h.

%%%%%%%%%%%%%%%%%%%%%%%%%%%%%%%%%%%%%%%%%%%%%%%%

% XMM Spectral Analysis

%%%%%%%%%%%%%%%%%%%%%%%%%%%%%%%%%%%%%%%%%%%%%%%%
\subsection{XMM Spectral Analysis}
\label{sec:spectral}

We performed a preliminary, model independent check for spectral variability by calculating the hardness ratio, defined as
%%%%%%%%%% Equation: Hardness Ratio
\begin{equation}
	HR = \frac{H-S}{H+S},
\label{eqn:HR}
\end{equation}
%%%%%%%%%% End: Equation 
{\noindent where $H$ is the counts in the hard X-ray band (5--10 keV in this work) and $S$ is the counts in the soft X-ray band(0.2--5 keV). The calculated hardness ratio between these two bands is shown in Fig. \ref{fig:HRvsT} with 400\,s binning. Given the evident evolution in hardness ratio, we extracted spectra from temporal regions showing similar flux behaviour and constant hardness ratio. The time intervals associated with these extraction regions are listed in Table \ref{tab:Tstamps} and correspond to the regions specified by dot--dashed lines in Fig. \ref{fig:LCwHR}. Region A corresponds to the low flux, early period of the observation. Regions B, C, D, G and H correspond to small flares that occur after the initial low flux interval but either side of the main flare period. Data from these flares were combined to improve the statistical quality of the data products. The mean values of hardness ratio for each of these regions is plotted in Fig. \ref{fig:LCwHR} as red circles. The hardness ratio was also calculated for the periods of intermediate behaviour not covered by regions A--H and is taken to represent the average behaviour of the source. This average hardness ratio is plotted as the red dashed line in Fig. \ref{fig:LCwHR}. The hardness ratios calculated for regions A--H show evidence of increasing from a value of approximately 0.25 to 0.7, coinciding with the rise to the main flare (regions E+F). 

%%%%%%%%%% Figure:HR vs Flux
\begin{figure}
\begin{center}
\includegraphics[width=0.5\textwidth,natwidth=576,natheight=432]{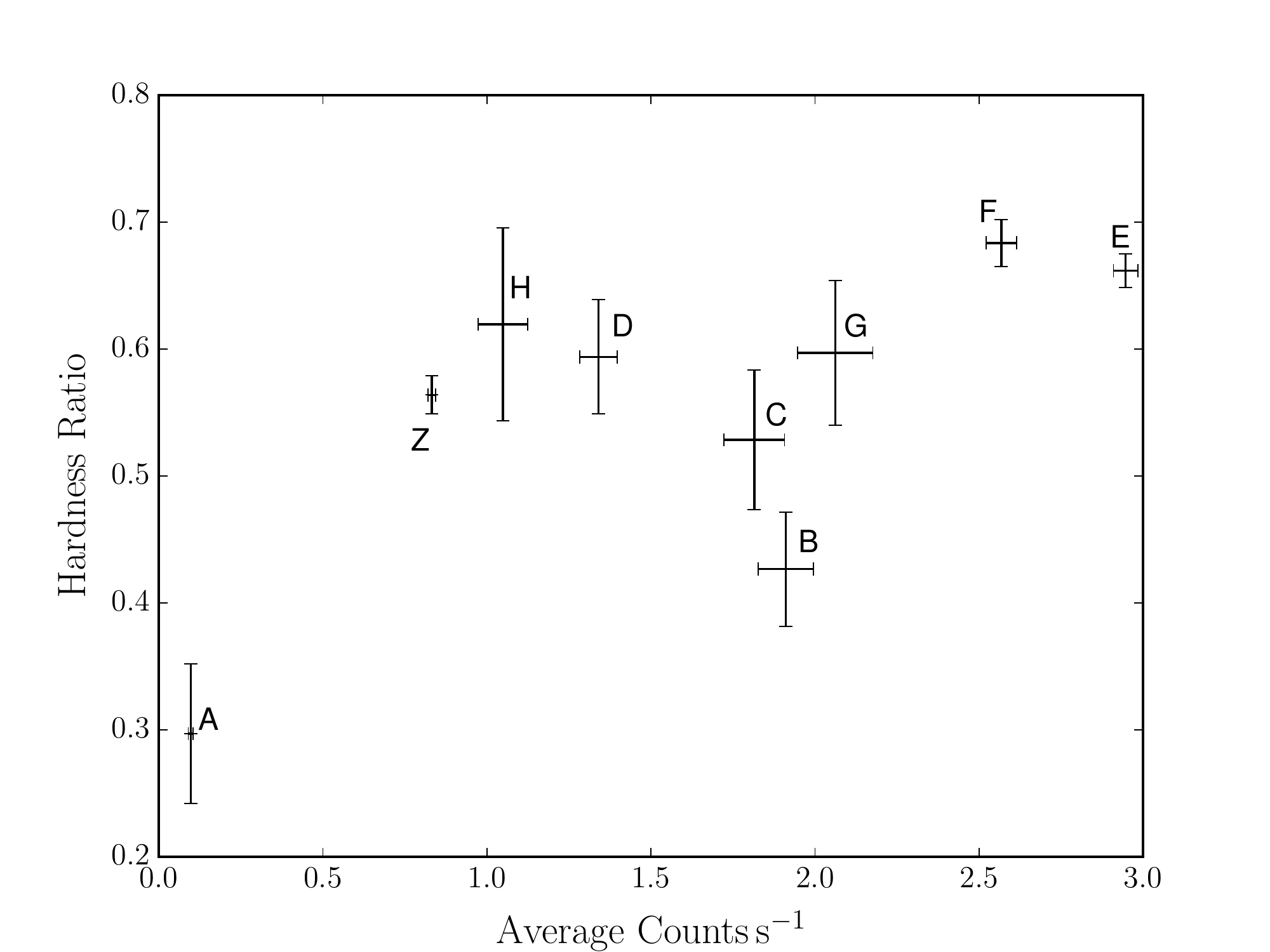}
\caption{ Hardness ratios taken from regions shown in Fig. \ref{fig:LCwHR} plotted against average counts rate for those regions. The point labelled `Z' corresponds to the hardness ratio and average flux for the time intervals not covered the regions labelled A--H.} 
\label{fig:HRvsFlux}
\end{center}
\end{figure}
%%%%%%%%%% End Figure

Fig. \ref{fig:HRvsFlux} shows the hardness ratio calculated for the spectral extraction regions plotted against average flux for these regions. This shows the often seen `harder-when-brighter' relation in HMXBs (\citealt{2012MNRAS.420..554S}; \citealt{2013MNRAS.429.2763S}; \citealt{2013ApJ...767...70O}; \citealt{2014A&A...562A...2R}; \citealt{2014ATel.5980....1K}) with spectra A, E and F clearly showing this behaviour. However, in the intermediate intensity range, the situation is somewhat less clear. We note that although this relation is often seen in HXMBs, the observed variation in this source is an extreme example of this behaviour.

In order to investigate the hardening of the emission from the source EPIC-pn and EPIC-MOS spectra from the regions shown in Fig. \ref{fig:LCwHR} and Table \ref{tab:Tstamps} were extracted with a minimum of 15 counts per bin. The pn and MOS spectra were fit simultaneously for each region using {\scshape xspec} in the 0.5--15\,keV energy band. In these cases, in order to get values of unabsorbed flux, the models {\it cons*phabs*cflux(powerlaw)} and {\it cons*phabs*cflux(bbody)} were fit to all spectra, where the constant here is used to account for the difference between the MOS and pn instruments. The model component {\it cflux} is a convolution model used to determine the 0.5--10\,keV flux of the model it is used in. The position of this component in this model gives the 0.5--10\,keV unabsorbed flux. Blackbody fits to the spectra gave unphysical blackbody temperatures in the range of 2--4\,keV for regions E, F and B+C+D+G+H. The fits of the power-law model to the spectra are presented in Table \ref{tab:ParamFits} and the power-law fits to regions A, B+C+D+G+H and E are shown in Fig. \ref{fig:RegSpectra}. The spectrum of region F is similar to that of region E and hence is not shown in Fig. \ref{fig:RegSpectra}.

The power-law model shows variation in both absorbing column density and photon index coinciding with the main period of flaring activity. The variation in photon index covers nearly an order of magnitude from $\sim$2.5 at early times in the observation to $\sim$0.36 at the time of the main flaring activity. These variations towards a flatter power-law are consistent with the evolution of the hardness ratio shown in Fig. \ref{fig:LCwHR}. 

More physical models often used in fitting SFXT spectra such as a Comptonizing plasma model [{\it phabs(compTT)}] or a cut-off power law [{\it phabs(cutoffpl)})] were also fit to the spectra of the regions listed in Table \ref{tab:ParamFits}. However, both of these alternative models did not give better fits to the data. Even in the case of a comparable fit, model parameters such as high-energy cut off or seed photon and plasma temperature were highly unconstrained. For example, for region E, the Comptonizing plasma model fit has a reduced chi squared, $\chi^{2}_{\mathrm{red}} =1.51$ (300 dof) with $kT = 2\pm 31\, \mathrm{keV}$, seed photon temperature $T_0 = 3.8\pm1.6 \, \mathrm{keV}$ and optical depth $\tau = 0.01\pm4$. For the same region, the cut off power-law fit gives the same absorbing column density and photon index as the power-law fit, however the high energy cut off is unconstrained ($500\pm10,000\, \mathrm{keV}$). 
}

%%%%%%%%%% Table: Spectrum Timestamps
\begin{table}
\begin{center}
\caption{Times associated with the regions used for spectral and hardness ratio analysis. T$_{\mathrm{start}}$ and T$_{\mathrm{stop}}$ are given in seconds since beginning of the observation.}
\begin{tabular}{ccccc}
\hline
\hline
Spectrum & T$_{\mathrm{start}}$ &T$_{\mathrm{stop}}$ & MJD$_{\mathrm{start}}$ & MJD$_{\mathrm{stop}}$\\
\hline
A & 0     & 6000   & 56372.530 & 56372.599\\
B & 8400  & 8800   & 56372.627 & 56372.632\\
C & 10200 & 10500  & 56372.648 & 56372.652\\
D & 12600 & 13300  & 56372.676 & 56372.684\\
E & 15800 & 20500  & 56372.713 & 56372.767\\
F & 20500 & 23800  & 56372.767 & 56372.806\\
G & 25500 & 25800  & 56372.825 & 56372.829\\
H & 27100 & 27500  & 56372.844 & 56372.848\\
\hline
\hline
\end{tabular}
\label{tab:Tstamps}
\end{center}
\end{table}
%%%%%%%%%% End: Table

Due to the unphysical nature of the blackbody temperature in the fits to regions E, F and B+C+D+G+H, we use the power-law fits throughout to characterise the spectral variability observed over the course of the observation. 

%%%%%%%%%% Figure: Spectra of Region Pre6000, Flare 3 and Rise
\begin{figure}
\begin{center}
\includegraphics[width=0.52\textwidth,natwidth=790,natheight=498]{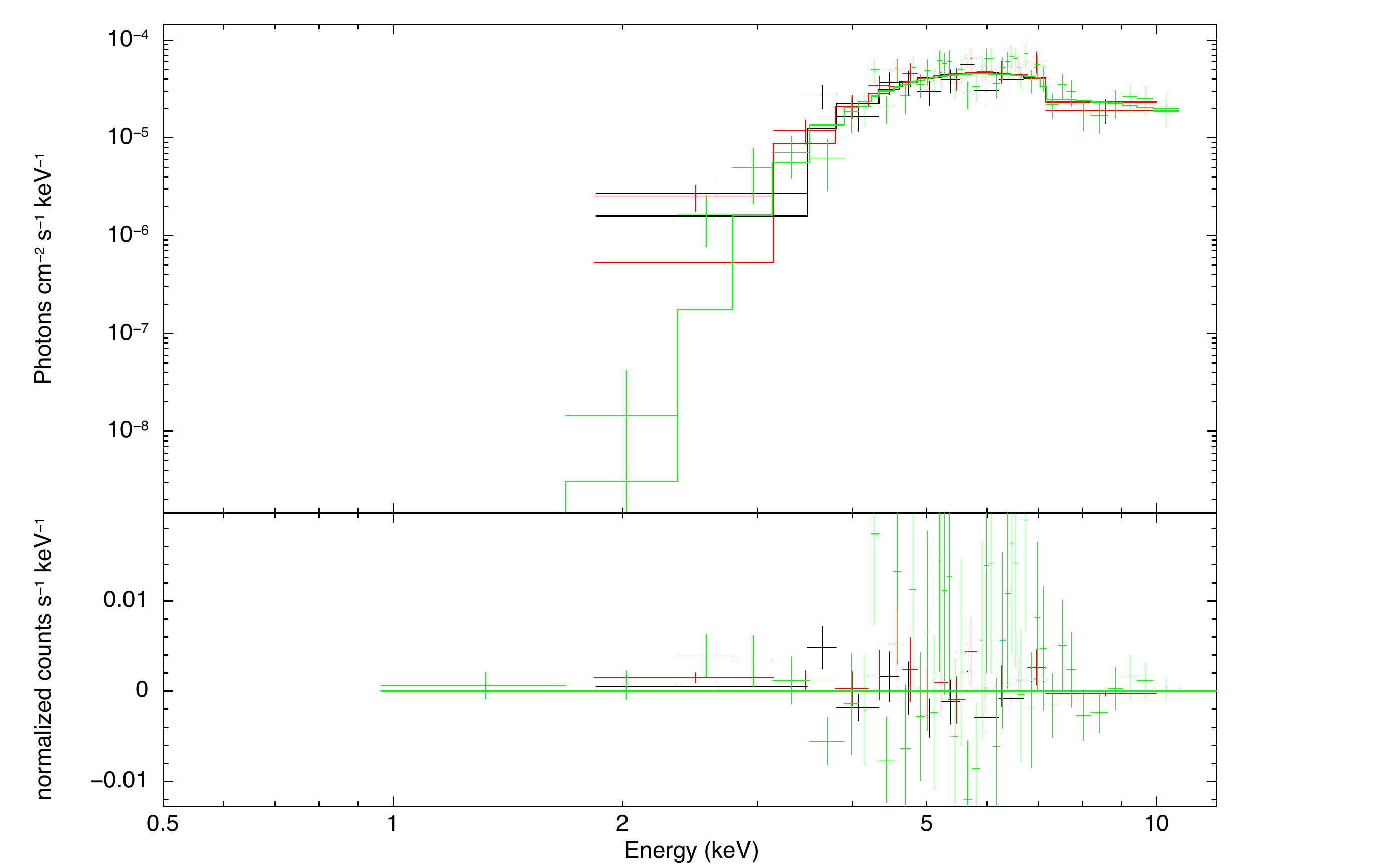}
\includegraphics[width=0.52\textwidth,natwidth=792,natheight=511]{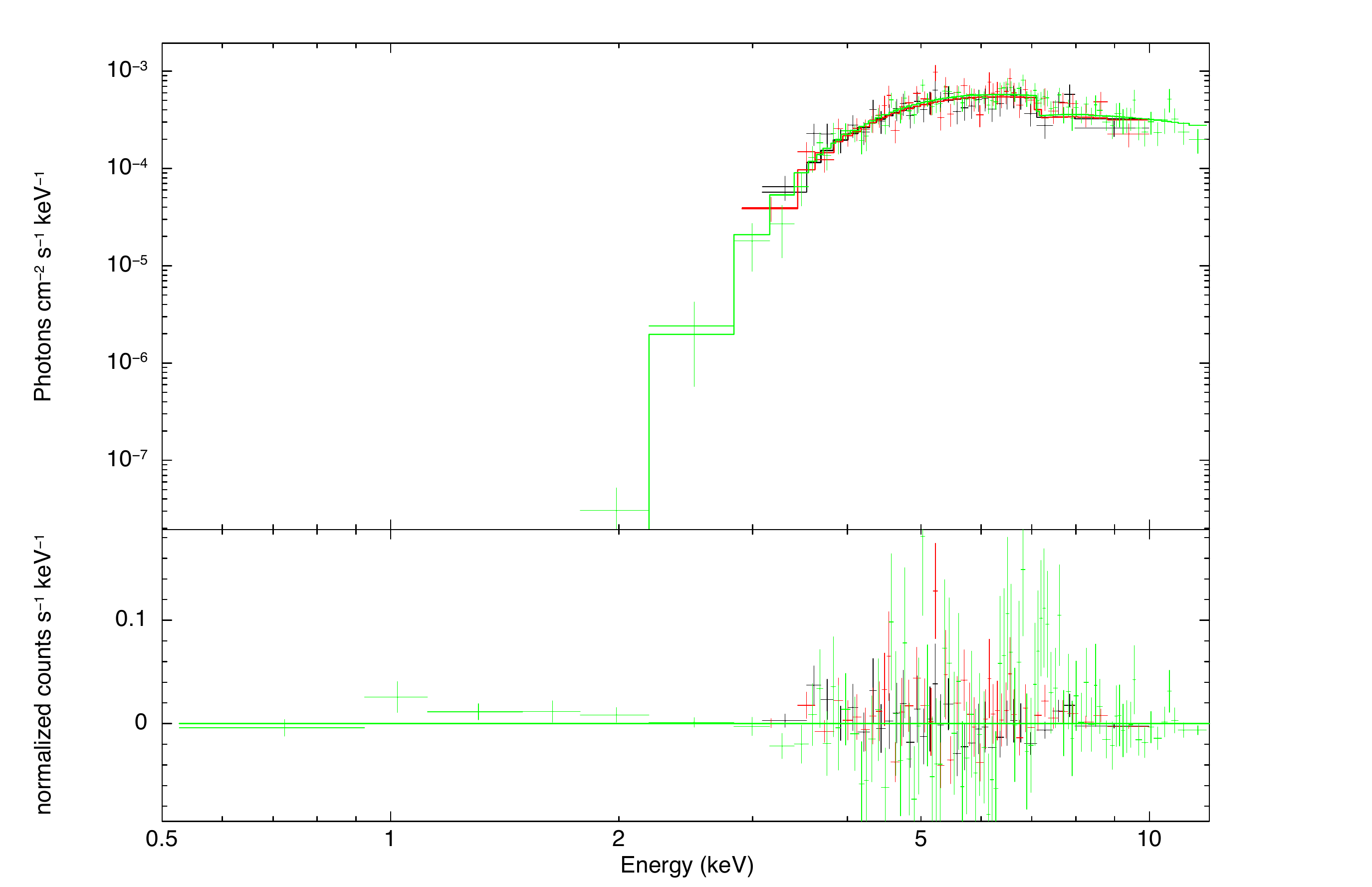}
\includegraphics[width=0.52\textwidth,natwidth=791,natheight=512]{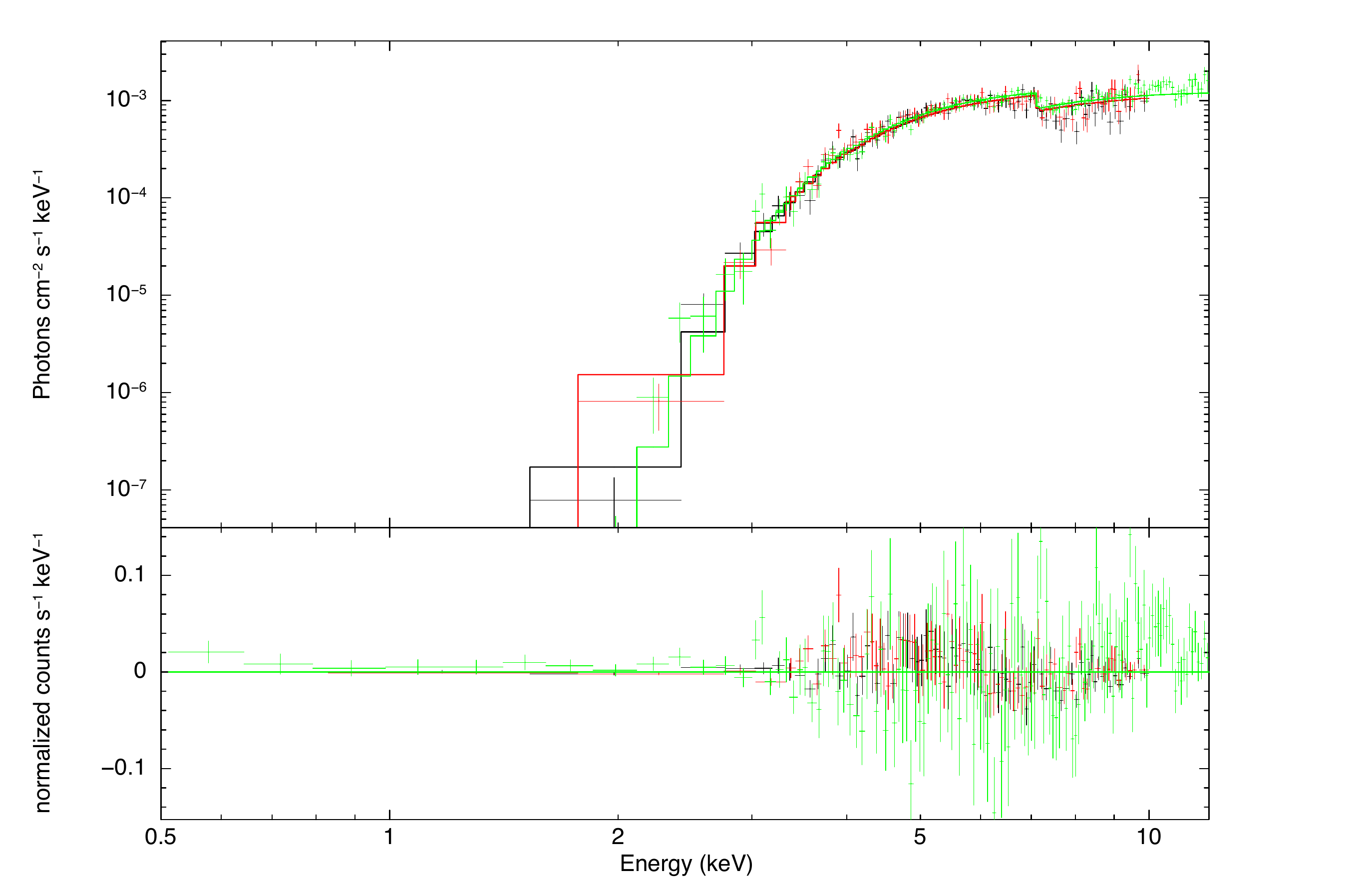}
\caption{Top Panel: power-law fit and residuals to EMOS (Black and Red) and EPN (Green) spectra from region A. Middle Panel: power-law fit and residuals to EMOS (black and red) and EPN (green) spectra from Region B+C+D+G+H. Bottom Panel: power-law fit and residuals to EMOS (black and red) and EPN (green) spectra from region E.}
\label{fig:RegSpectra}
\end{center}
\end{figure} 
%%%%%%%%%% End: Figure 

\begin{figure}
\begin{center}
\includegraphics[width=0.5\textwidth,natwidth=784,natheight=551]{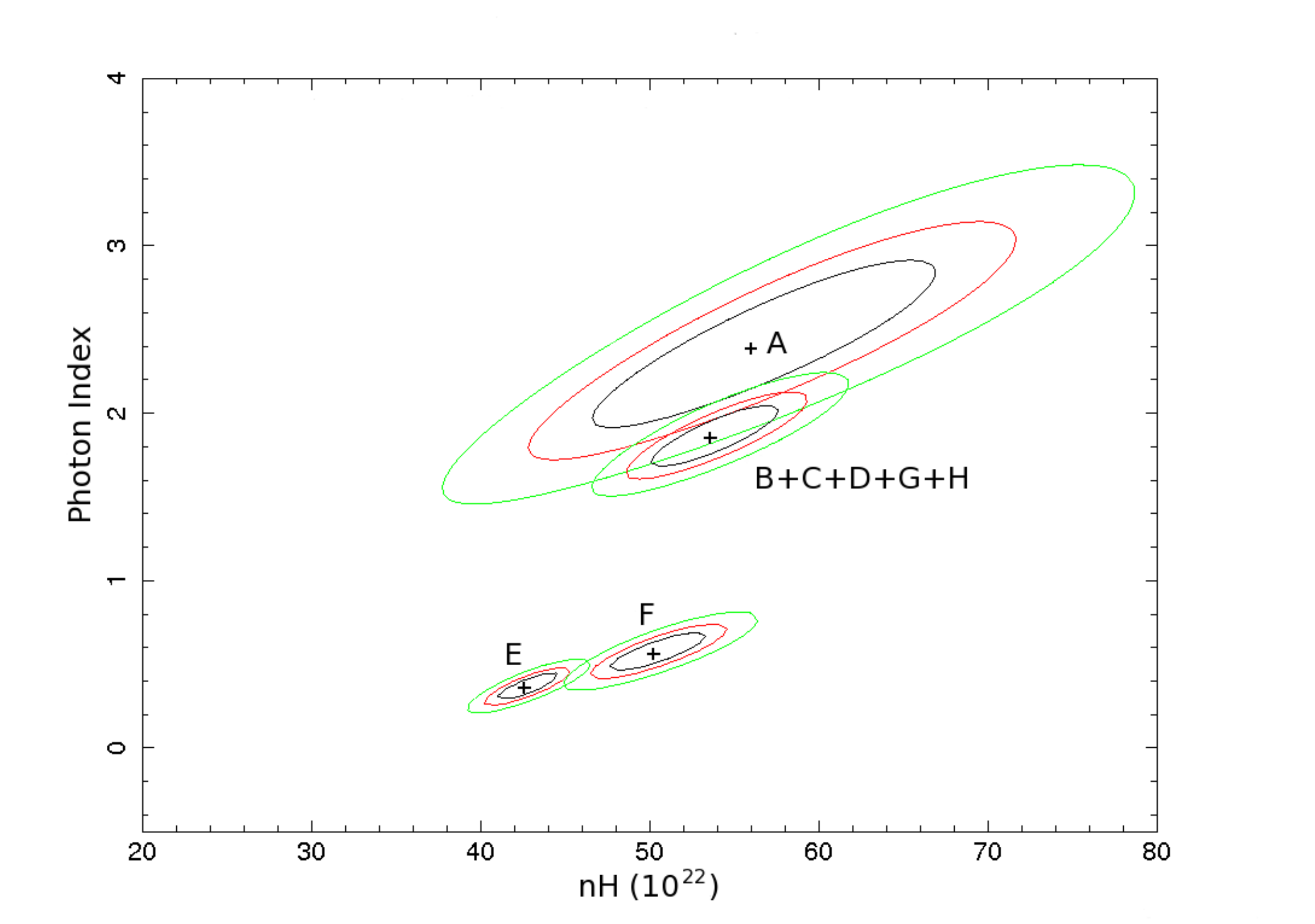}
\caption{Confidence contours for the best-fitting spectral parameters to regions A, B+C+D+G+H, E and F using an absorbed power law model [{\it cons*phabs*cflux(powerlaw)}] with the 68\%, 90\% and 99\% statistical confidence regions shown. }
\label{fig:Contours}
\end{center}
\end{figure}

The best-fitting values of photon index show significant evidence for spectral evolution over the observation. The 68\%, 90\% and 99\% statistical confidence regions for these two parameters in regions A, B+C+D+G+H, E and F are shown in Fig. \ref{fig:Contours} and show the evolution of the photon index over the course of the observation. It is clear that the spectra of the low-flux state (region A) and the main flaring activity (regions E+F) occupy distinct regions of the parameter space. By combining data from all the short flares, we can also see that their spectrum is different to regions A, E and F.

%%%%%%%%%% Figure: Contour plot of Spectra
%%%%%%%%%% End: Figure 

\begin{table*}
\begin{center}
\caption{Hardness ratios and best-fitting parameters for a power-law model [{\it cons*phabs*cflux(powerlaw)}] fit to spectra extracted from the regions outlined in Fig. \ref{fig:LCwHR} and Table \ref{tab:Tstamps}. Absorbing column density, n$_{H}$ is in units of  10$^{22}$\,cm$^{-2}$, the 0.5--10\,keV unabsorbed flux is in units of  $\times$\,10$^{-10} \, \mathrm{erg\,cm^{-2}\,s^{-1}}$.}
\begin{tabular}{|c|c|c|c|c|}
\hline
\hline
 Parameter & A & E & F & B+C+D+G+H  \\
\hline
Hardness Ratio & 0.30$\pm$0.06 & 0.66$\pm$0.01 & 0.68$\pm$0.02 & 0.54$\pm$0.02  \\  
\hline
n$_{H}$  & 56$^{+12}_{-10}$   & 43$^{+2}_{-2}$ & 51$^{+3}_{-3}$ & 53$^{+4}_{-4}$ \\[4pt]
$\Gamma$  & 2.4$^{+0.6}_{-0.5}$ & 0.36$^{+0.1}_{-0.1}$ & 0.57$^{+0.1}_{-0.1}$ & 1.86$\pm$0.2\\[4pt]
Unabsorbed Flux              & 0.27$^{+0.38}_{-0.13}$ &  1.49$^{+0.07}_{-0.07}$ & 1.66$^{+0.16}_{-0.16}$ & 1.82$^{+0.50}_{-0.37}$\\[4pt]
$\chi^{2}_{red}$ ({\it dof}) & 1.07 (68) &  1.34 (302) & 1.39 (242) & 1.02 (163)\\[4pt]

\hline
\hline

\end{tabular}
\label{tab:ParamFits}
\end{center}
\end{table*}
The source shows variation in photon index and when comparing this to the light curve in Fig. \ref{fig:LCwHR}, the source evolves from a soft photon index at the onset of the flaring behaviour to hard photon index around the main flare event. Due to the statistical properties of the spectra, we are unable to investigate the applicability of more complex models in regions E and F by subdividing these regions into smaller intervals

%%%%%%%%%%%%%%%%%%%%%%%%%%%%%%%%%%%%%%%%%%%%%%%%

% Combined XMM & INTEGRAL spectral

%%%%%%%%%%%%%%%%%%%%%%%%%%%%%%%%%%%%%%%%%%%%%%%%
\subsection{\emph{INTEGRAL} and Combined Spectral Analysis}

With the aim of better constraining the spectral model of the source, an average \emph{INTEGRAL}/IBIS spectrum was extracted from the ScWs with the greatest significance detections of the source in the first observation in Revolution 1274. This spectrum was fitted in the 18--60\,keV range with a power-law model, yielded a photon index of $\Gamma\,=\,3.15^{+1.15}_{-0.99}$ with $\chi^{2}_{\mathrm{red}}\,=\,0.80$ (2 {\it dof}) and a model flux of $2.879\times10^{-10}$ \lumin. 

Simultaneous fits of the \emph{INTEGRAL} spectrum with the \emph{XMM} MOS and pn spectra were performed in the 0.5--50\,keV range. Fits of an absorbed power-law model described in Section \ref{sec:spectral} yielded good fits and best-fitting parameters consistent with those presented in Table \ref{tab:ParamFits} for the \emph{XMM} data only. The statistics of the \emph{INTEGRAL} spectrum are such that they do not constrain the hard X-ray spectral shape well and consequently the much better statistics in the \emph{XMM} energy range cause it to dominate when fitting over the full spectral range. As a result, combining \emph{XMM} and \emph{INTEGRAL} spectra does not allow us to better constrain the spectral model achieved with \emph{XMM} data alone. 
%%%%%%%%%%%%%%%%%%%%%%%%%%%%%%%%%%%%%%%%%%%%%%%%%
%
%% Discussion
%
%%%%%%%%%%%%%%%%%%%%%%%%%%%%%%%%%%%%%%%%%%%%%%%%%
\section{Discussion}
\label{sec:Discussion}
We present the results of a 30\,ks \emph{XMM--Newton} observation of the SFXT, SAX J1818.6$-$1703. This is the first in-depth soft X-ray investigation of the source located around the peak in the \emph{INTEGRAL/IBIS} phase folded light curve (fig. 3 of \citealt{Bird2008}), likely located near periastron. We also present the results of contemporaneous archival \emph{INTEGRAL}/IBIS observations spanning 115\,ks with 25\,ks on-source exposure. 

\subsection{\emph{INTEGRAL} Context}
\label{sec:INT_context}
Previous studies of this source with \emph{INTEGRAL} have revealed a long (30\,d), eccentric (e\,$\sim$\,0.3\,--\,0.4) orbit and shown detections in more than 60\% of neutron star periastron passages. 

SAX J1818.6$-$1703 was not detected by \emph{INTEGRAL} in an observation carried out between 2013 March 12 and 2013 March 13 during Revolution 1271, at a time that corresponds to an orbital phase of 0.804\,--\,0.862. During this region of a long eccentric orbit, the stellar wind, through which the neutron star passes, is not expected to be sufficiently dense or structured so as to produce detectable emission from accretion. 

A bright flare from SAX J1818.6$-$1703 detected by {\it Swift}/BAT on 2013 March 17 (corresponding to phase, $\phi$\,$=$\,0.970) was reported by \citet{2014A&A...562A...2R}. During this flare the source reached a peak flux of 146 mCrab (15--50\,keV), clearly showing that the source exhibited enhanced X-ray emission a few days prior to the \emph{XMM} observation reported here. 

The source was also detected by \emph{INTEGRAL} during Revolution 1274 at a significance of $\sim$13$\sigma$. A more detailed analysis of this revolution shows that the source was first observed between UTC 11:10:03 and UTC 20:44:48 on 2013 March 20, corresponding to orbital phases $\phi\,=\,$0.086\,--\,0.099, with a total source exposure of 7.5\,ks. During this observation, the source was detected with an average count rate of 2.17$\pm$0.22 counts s$^{-1}$ (18--60 keV) or $\sim$10\,mCrab. The source was subsequently observed between UTC 09:00:11 and UTC 20:13:27 on 2013 March 22, corresponding to orbital phases $\phi\,=\,$0.150\,--\,0.165, again with a total source exposure of 7.5\,ks. The average count rate in this observation was 1.27$\pm$0.22 counts s$^{-1}$ (18--60 keV). In order to compare these fluxes to previous periastron passages, we performed the recurrence analysis of \citet{Bird2008} using an \emph{INTEGRAL}/IBIS light curve of SAX J1818.6$-$1703 from the first 1000 revolutions of \emph{INTEGRAL}. We calculated the mean flux detected by \emph{INTEGRAL} during a 4-day period centred on periastron and from this, we find that during the first observation of Revolution 1274, the source is brighter than 80\% of periastron passages observed by \emph{INTEGRAL}. The flux detected between UTC 09:00:11 and UTC 20:13:27 on 2013 March 22 is consistent with fluxes observed by \emph{INTEGRAL} during the majority of periastron passages. However, it is different from the distribution of apastron flux measurements and still represents a significant detection of the source. When considering both the {\it Swift}/BAT bright flare on 2013 March 17 and the \emph{INTEGRAL} observations during Revolution 1274, it is clear that SAX J1818.6$-$1703 was in an atypically active state during the periastron passage covered by the \emph{INTEGRAl} and \emph{XMM} observations presented in this work.

\subsection{Source Distance}
\label{sec:Distance}
Knowledge of the source distance is critical in order to calculate associated luminosities and hence disentangle the accretion processes generating the observed dynamic X-ray behaviour. A number of source distances have been suggested for SAX J1818.6$-$1703, though the distance is still one of the best constrained for an SFXT. Using near infra-red spectroscopy \cite{2010A&A...510A..61T} inferred a distance of 2.1$\pm$0.1\,kpc. More recently, \cite{2013ApJ...764..185C} inferred a distance of 2.7$\pm$0.28 using a similar method. For the subsequent discussion, we adopt the better constrained distance estimate of 2.1$\pm$0.1\,kpc from \cite{2010A&A...510A..61T}. The range of possible distances to the source is such that the luminosities calculated assuming a distance of 2.1\,kpc could be a factor of 1.3 larger if the source is located at 2.7\,kpc.

\subsection{Observed Luminosity}
\label{sec:ObsLumin}
\begin{figure}
\begin{center}
\includegraphics[width=0.5\textwidth,natwidth=576,natheight=432]{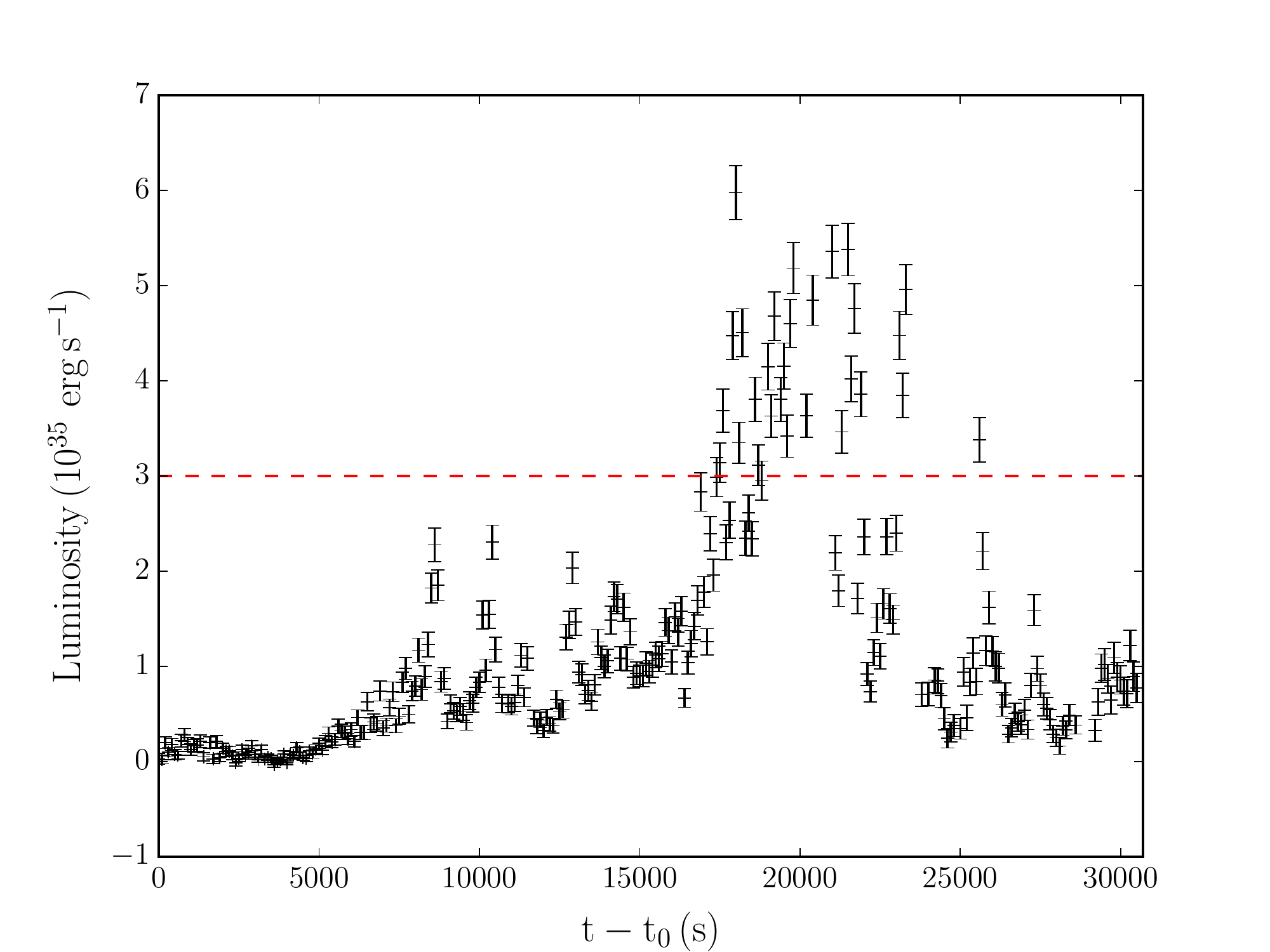}
\caption{The \emph{XMM--Newton} light curve shown in Fig. \ref{fig:LCwHR} with count rate converted into luminosity using an absorbed power law with photon index $\Gamma\,=\,1.9$, absorbing column density, n$_{H}\,=\,5\times10^{23}\, \mathrm{cm^{-2}}$ and a source distance of $2.1\pm0.1$\,kpc. The red dashed line corresponds to the critical luminosity for transition between the radiatively cooled to Compton-cooled regimes in \citet{2013MNRAS.428..670S}}
\label{fig:LuminLC}
\end{center}
\end{figure}

In order to critically evaluate the accretion processes taking place over the course of the \emph{XMM--Newton} observation, the pn count rates shown in Fig. \ref{fig:LCwHR} were converted to X-ray luminosities. To convert from count rate to luminosity, we assumed an absorbed power-law spectral model with photon index $\Gamma\,=\,1.9$, an absorbing column density of n$_{H} \,=\, 5\times10^{23}\, \mathrm{cm^{-2}}$ and a source distance of 2.1$\pm$0.1\,kpc. These spectral parameters are consistent with the fits to regions A and B+C+D+G+H. Although these are not consistent with regions E and F, we apply the same model to these sections of the light curve so as not to introduce discontinuities in the luminosity light curve scale factor. The resulting luminosity light curve is shown in Fig. \ref{fig:LuminLC}. 

During the first 6\,ks of the \emph{XMM} observation, the source exhibits low level persistent emission with luminosities of $\sim10^{34}$ \lumin. This corresponds to low level accretion of material from the stellar wind, however the light curve shows no flaring behaviour during this interval. We investigate this further by estimating the 0.5--10\,keV luminosity for the \emph{INTEGRAL} observation of the source between UTC 11:10:03 and UTC 20:44:48 on 2013 March 20. Using an \emph{INTEGRAL}/IBIS source count rate of 2.17 counts s$^{-1}$ and the cut-off power law of \cite{2009MNRAS.400..258S} and the absorbing column density measured in this work of $5\times10^{23}\, \mathrm{cm^{-2}}$, we calculate an \emph{XMM} count rate of 4.45 counts s$^{-1}$ (0.5--10\,keV). Assuming the same absorbed power-law used to produce Fig. \ref{fig:LuminLC}, this corresponds to a luminosity of $3.3\times10^{35}$\,\lumin. This suggests that the initial 6\,ks of the \emph{XMM} observation is a low flux interval between periods of higher activity. 

One possible explanation for initial low flux observed by \emph{XMM--Newton} could be the obscuration of the emitting region by circumstellar material. However, the spectral analysis of Section \ref{sec:spectral} rules out this scenario as the values of absorbing column density for this region are consistent with the regions showing flaring activity. 

Excluding the obscuration of the source by a dense clump, other possibilities are often discussed in the literature to explain these `dips' or `off-states' sometimes observed in supergiant HMXB pulsars. These include a transition to a less effective accretion regime likely triggered by a mild variability in the wind properties. If this is the case, the initial dip could be explained by either a transition to the subsonic propeller regime where the low luminosity is produced by means of matter penetrating the magnetosphere by the Kelvin-Helmholtz instability \citep{2011A&A...529A..52D} or by a cooling regime switch \citep{2013MNRAS.428..670S}. In this latter scenario, an accretion regime transition from a Compton cooling dominated regime (implying a higher accretion rate through the magnetosphere) to a radiative dominated regime (producing a lower X-ray luminosity) is produced in the equatorial plane of the neutron star magnetosphere, due to a switch from a fan beam to a pencil beam geometry \citep{2013MNRAS.428..670S}.

Following the method of \cite{2011A&A...529A..52D} and \cite{2013MNRAS.433..528D}, we apply the framework of \cite{2008ApJ...683.1031B} to explain this low activity region seen in the first 6\,ks of the \emph{XMM} observation. For supergiant stellar parameters, we adopt a stellar mass and radius of 25\,M$_{\odot}$ and 30\,R$_{\odot}$ respectively, a stellar mass loss rate $\dot{M}_{w}\,=\,5\times10^{-7}$\,M$_{\odot}$\,yr$^{-1}$ and terminal wind velocity of 1000\,km s$^{-1}$. We also assume a magnetic field strength of $10^{12}\,\mathrm{G}$ for the neutron star, in line with measurements of other field strengths in HMXBs and the SFXT, IGR J17544$-$2619 (\citealt{2002A&A...395..129K}, \citealt{2015MNRAS.447.2274B}). Using an average luminosity of $1\times10^{34}$ \lumin, equation 21 of \cite{2008ApJ...683.1031B}, the range of wind velocities and plasma density ratios used by \cite{2013MNRAS.433..528D}, the corresponding neutron star spin periods fall between 50 and 90\,s. Hence the sub-sonic propeller regime could be invoked to explain this low activity period if the neutron star is slowly rotating with a magnetic field strength typical of HMXB pulsars.

Given the low luminosities associated with region A, we can also apply the quasi-spherical accretion theory of \cite{Shakura2012a}. In this framework, for luminosities below $\sim3\times10^{35}$\, \lumin, the source inhabits a region where plasma held above the magnetosphere cools by thermal bremsstrahlung before entering the magnetosphere by the Rayleigh Taylor instability \citep{2013MNRAS.428..670S}. The low luminosity interval at the beginning of the \emph{XMM} observation can be explained as lower density plasma entering the magnetosphere during this time. 

Recently, hydrodynamic simulations of Vela X-1, taking into account ionisation of the stellar wind by the central X-ray source, by \cite{2015A&A...575A..58M} have shown that off-states lasting between 300 and 7.2ks can be generated by lower density `bubble' structures forming in the wind around the neutron star. The density of these bubbles is approximately one tenth that of the time-averaged density of the plasma surrounding the neutron star causing a corresponding luminosity drop of the same factor. Therefore, the presence of one of these low density bubbles could also explain the first 6\,ks of this \emph{XMM} observation 

The increase in luminosity following the low activity region can also be explained within this framework as increases in the amount of material entering the magnetosphere while still in the radiative cooling regime. The main flare event (regions E and F) at approximately 17 500\,ks exceeds the critical luminosity value of $\sim3\times10^{35}$ \lumin (shown as a red dashed line in Fig. \ref{fig:LuminLC}) for a transition between the radiative and Compton cooling regimes \citep{2013MNRAS.428..670S}. This occurs at approximately the same time as the evolution to a harder photon index in the spectral fits. 

After the period of enhanced soft X-ray emission, the source decreases in luminosity and appears to return to the radiatively cooled regime, again with flaring behaviour. Using the second \emph{INTEGRAL} observation of SAX J1818.6$-$1703 during Revolution 1274, we can infer the soft X-ray behaviour of the source after the \emph{XMM} observation. During the \emph{INTEGRAL} observation, an average source flux of 1.27 counts s$^{-1}$ was detected. Using the same method as for the previous \emph{INTEGRAL} observation between UTC 11:10:03 and UTC 20:44:48 on 2013 March 20, we infer an \emph{XMM} 0.5--10\,keV count rate of 2.6 counts s$^{-1}$. Assuming an absorbed power law with absorbing column density, n$_{H} \,=\, 5\times10^{23}\, \mathrm{cm^{-2}}$ and photon index, $\Gamma\,=\,1.9$, this corresponds to a luminosity of $1.9\times10^{35}$\,\lumin. This is consistent with continued flaring behaviour following the \emph{XMM--Newton} observation of SAX J1818.6$-$1703.

We note that although the distribution of orbital eccentricities of SFXTs is not well known (see \citealt{Paizis2014} and references therein), short-period systems are likely to have more circular orbits. The high eccentricity of SAX J1818.6$-$1703 and resulting wind density variation along the orbit experienced by the compact object is likely to be an additional factor and acting alongside the mechanisms discussed above driving the atypical variability seen in this source. 

\subsection{Spectral Variation}
\label{sec:Spectral Variation}

The analysis of the hardness ratio evolution shows that SAX~J1818.6$-$1703 undergoes clear spectral evolution, shown in Fig. \ref{fig:Contours}, during the development of the flaring behaviour. The `harder-when-brighter' relation often seen in HMXBs is also seen in this observation, as shown in Fig. \ref{fig:HRvsFlux}. Spectra A, E and F clearly show this behaviour, however in the intermediate intensity range, this relationship is less clear. One explanation for this could be that the average source count rate is not representative of source luminosity as the spectral extraction regions contain both flare peaks and lower intensity portions of the light curve. Changes in absorption caused by circumstellar material could also mask or distort the `harder-when-brighter' relation, but we see no evidence for that here. 

A more detailed analysis (Table \ref{tab:ParamFits}) revealed variations in the photon index. The source shows significant variation in photon index and when comparing this to the EPIC-pn light curve of Fig. \ref{fig:LCwHR}, it is clear that the source evolves from a soft spectrum at the onset of the flaring episode to a harder spectrum around the main flare event. From Fig. \ref{fig:Contours}, using a power law fit, it is evident that this is a true variation within this model and not a result of the correlation of the fit parameters as the parameters are seen to be only weakly correlated. However, using a more realistic absorption model could change the observed relationship between the absorption and spectral slope. Although the evolution shown in Fig. \ref{fig:Contours} is model-dependent, evolution in the hardness ratio shows that higher energy photons are being detected for periods of higher flux. An absorbed power law is one method of characterising this change which gives reasonable fits to the data.

It can be seen from Table \ref{tab:ParamFits} that the photon index of the underlying continuum varies from 0.3 to 2.5 on a time-scale of approximately 10\,ks. The absolute values of absorbing column density are among the highest ever observed in an SFXT and are comparable with the observed $(6\pm1)\times10^{23}\, \mathrm{cm^{-2}}$ in IGR J16465$-$4507 \citep{2006A&A...453..133W}. Previous soft X-ray observations of an outburst of this source around periastron by \cite{2009MNRAS.400..258S} with {\it Swift} did not reveal any spectral variability in either photon index or absorbing column density; however, the source was in outburst for the duration of the observation. Those observations did reveal a high absorbing column density of (5--7)$\times\,10^{22}\,\mathrm{cm^{-2}}$ and a photon index of $\Gamma\,=\,0.3\pm0.2$ for a cut-off power-law fit to {\it Swift}/XRT and BAT data. This absorption is still an order of magnitude less than that measured in this observation, while the photon index is consistent with that measured here. Such large values of absorbing column density are more in line with those measured in the highly obscured persistent HMXBs discovered by \emph{INTEGRAL} (\citealt{2003A&A...411L.427W}; \citealt{2004ApJ...616..469F}). Given the line of sight absorption to SAX J1818.6$-$1703 (n$_{H}\,=\,1.42\times10^{22}\,\mathrm{cm^{-2}}$; \citealt{1990ARA&A..28..215D}), it is clear that this increased absorption is intrinsic to the system. 

%%%%%%%%%% Figure:Luminosity LC
%%%%%%%%%% End Figure

%%%%%%%%%%%%%%%%%%%%%%%%%%%%%%%%%%%%%%%%%%%%%%%%%
%
%% Temporal Discussion
%
%%%%%%%%%%%%%%%%%%%%%%%%%%%%%%%%%%%%%%%%%%%%%%%%%
\subsection{Temporal Studies}
Spin periods of neutron stars in SFXTs are known for less than half of the known sources and range from $\sim$5\,s in AX J1841.0-0536 (\citealt{2001PASJ...53.1179B}, see also \citealt{2011MNRAS.412L..30R} for further discussion) to $\sim$1212\,s in the case of IGR J16418--4532 \citep{2012MNRAS.420..554S}. Previous observations of SAX J1818.6$-$1703 with \emph{XMM--Newton} and {\it Swift} have failed to find a periodic signal that could be interpreted as a neutron star spin period (e.g.\citealt{2009MNRAS.400..258S}; \citealt{2012A&A...544A.118B}). 

Searching for a neutron star spin period in this system using a both a Lomb--Scargle periodogram and an epoch folding method as described in Section \ref{sec:XMMtemporal}, we detected no significant periodicities in the range of 2\,s--4.2\,h. This period range is the range over which the data set would allow for a periodic signal to be detected with confidence and covers the spread in known neutron star spin periods in SFXTs \citep{Paizis2014}. Since spin periods in some systems have only been detected in certain accretion states, we selected time windows corresponding to periods of lower and higher intensity from the EPIC-PN light curve (t $<$ 8500, 8500 $<$ t $<$ 15\,000, 15\,000 $<$ t $<$ 25\,000, t $>$ 25\,000) and tested for the presence of neutron star spin periods; however, no significant signal was detected. 

A possible explanation for the lack of a neutron star spin period detection comes from the nature of the behaviour during this observation. The source exhibits flaring activity, with 5 flares each lasting $\sim$1\,ks and a main flare event, seen to occur approximately 17.5\,ks after the start of the observation. The rapidly changing nature of the flux could have masked any periodic modulation due to the rotation of the neutron star.

Another explanation for the non-detection could be that the spin period value lies outside the range probed in this work. This would imply the neutron star is either a faster or slower rotator compared to other neutron stars hosted in SFXTs. Currently, the longest known neutron star spin period in an HMXB is $\sim$2.8\,h found in 2S 0114$+$650 \citep{1999ApJ...511..876C}. If the neutron star in SAX J1818.6$-$1703 were to have such a slowly rotating neutron star, it would be located at the extreme end of the spin period distribution for HMXBs (fig. 1 of \citealt{2014ApJ...786..127E};catalogue: \citealt{2006A&A...455.1165L}). 

One further possibility to explain the lack of pulsations would be a geometric effect such as low inclination of the binary plane or the existence of a small angle between rotation and magnetic field axes as suggested by \cite{1999A&A...349..873R} for the HMXB 4U 1700$-$37. 

The non-detection of a spin period allows the possibility that the compact object could be a black hole. However, there are currently no candidate black hole systems amongst SFXTs and only a handful in the entire population of HMXBs e.g Cyg X$-$1, LMC X$-$3 and MWC 656 (\citealt{1975ApJ...200..269B}, \citealt{1983ApJ...272..118C}, \citealt{2014Natur.505..378C}).

There is a hint of ``quasi-periodic'' flaring in the light curve, on a time-scale of about 2 ks. This could be explained by a number of models of HMXB accretion (see for example, the discussion in \citealt{2012MNRAS.420..554S} regarding quasi-periodicity in X-ray flaring of the SFXT, IGR J16418--4532).
The explanations that seem to better apply to SAX 1818.6$-$1703 include the quasi-spherical settling accretion model \citep{Shakura2012a}, where convective motion at the base of the shell located above the magnetosphere can produce X-ray flares on a quasi-periodic time scale, or the effects seen in hydrodynamic simulations of the accretion flow in the vicinity of the neutron star (\citealt{2015A&A...575A..58M} and references therein), where low-density bubbles forms and are accreted by the neutron star in a sort of `breathing mechanism', alternating between low- and higher density matter.

%%%%%%%%%%%%%%%%%%%%%%%%%%%%%%%%%%%%%%%%%%%%%%%%%
%
%% Conclusion
%
%%%%%%%%%%%%%%%%%%%%%%%%%%%%%%%%%%%%%%%%%%%%%%%%%
\section{Conclusions}
\label{sec:Conclusion}

SAX J1818.6$-$1703 has been observed, for the first time in soft X-rays, around periastron. The observed luminosities seen in both \emph{INTEGRAL} and \emph{XMM--Newton} place this source in a highly active state with periods of fast flaring followed by a longer duration flare event reaching a peak luminosity of $\sim6\times10^{35}$\, \lumin. The large increase in activity is coincident with a significant hardening of the source spectrum from a photon index $\Gamma\,\sim\,1.9$ to $\Gamma\,\sim\,0.3$, suggesting a change in accretion mechanism compared to the earlier flaring activity. Although other mechanisms cannot be ruled out, we note that the luminosity of the main flare period exceeds the critical luminosity required for a transition between the radiative and Compton cooling regimes in the theory of quasi-spherical accretion proposed by \citealt{Shakura2012a}. We also report an initial lower luminosity state with L$_{\mathrm{X}}\,\sim\,10^{34}$ \lumin that could be generated either by the source entering a subsonic propeller regime as outlined by \citet{2008ApJ...683.1031B} or by a decrease in density of material accreting on to the neutron star.

Spectral analysis indicates high intrinsic absorbing column densities of n$_{H} \, \sim \, 5 \times 10^{23}$ cm$^{-2}$; an order of magnitude higher than previous observations of the source. These column densities are amongst the highest measured in an SFXT and comparable to those found in highly obscured HMXBs discovered by \emph{INTEGRAL} \citep{2006A&A...453..133W}.

Further detailed investigation of this source relies upon a determination of the neutron star characteristics, principally its spin period and magnetic field strength, as has recently been demonstrated by \textit{NuSTAR} observation of IGR J17544$-$2619 \citep{2015MNRAS.447.2274B}. We therefore encourage further observations of SAX J1818.6$-$1703 with these aims in mind. 
%%%%%%%%%%%%%%%%%%%%%%%%%%%%%%%%%%%%%%%%%%%%%%%%

% Acknowledgments

%%%%%%%%%%%%%%%%%%%%%%%%%%%%%%%%%%%%%%%%%%%%%%%%
\section*{Acknowledgements}
The authors thank the referee, Roland Walter, for his useful comments and suggestions to significantly improve the quality of this work. This work was based on observations obtained with \emph{XMM–-Newton}, an ESA science mission with instruments and contributions directly funded by ESA Member States and NASA. CMB is supported by the UK Science and Technology Facilities Council. ABH acknowledges that this research was supported by a Marie Curie International Outgoing Fellowship within the 7th European Community Framework Programme (FP7/2007–2013) under grant no. 275861. LS and VS acknowledge the grant from PRIN-INAF 2014, ``Towards a unified picture of accretion in High Mass X-Ray Binaries''. MEG is supported by a Mayflower scholarship from the University of Southampton. SPD acknowledges support from the UK Science and Technology Facilities Council. This research has made use of the SIMBAD data base, operated at CDS, Strasbourg, France.
%%%%%%%%%%%%%%%%%%%%%%%%%%%%%%%%%%%%%%%%%%%%%%%%

% Bibliography

%%%%%%%%%%%%%%%%%%%%%%%%%%%%%%%%%%%%%%%%%%%%%%%%
%\newpage
\bibliography{Master_references.bib}

\begin{thebibliography}{63}
\providecommand{\natexlab}[1]{#1}

\bibitem[{{Arnaud}(1996)}]{1996ASPC..101...17A}
{Arnaud} K.~A., 1996, in G.H. {Jacoby}, J.~{Barnes}, eds, \aspc. Astronomical
  Society of the Pacific Conference Series, Vol. 101, p.~17

\bibitem[{{Bamba} et~al.(2001){Bamba}, {Yokogawa}, {Ueno}, {Koyama} \&
  {Yamauchi}}]{2001PASJ...53.1179B}
{Bamba} A., {Yokogawa} J., {Ueno} M., {Koyama} K., {Yamauchi} S., 2001, \pasj,
  53, 1179

\bibitem[{{Bhalerao} et~al.(2015)}]{2015MNRAS.447.2274B}
{Bhalerao} V. et~al., 2015, \mnras, 447, 2274

\bibitem[{Bird et~al.(2008)Bird, Bazzano, Hill, McBride, Sguera, Shaw \&
  Watkins}]{Bird2008}
Bird A.~J., Bazzano A., Hill A.~B., McBride V.~A., Sguera V., Shaw S.~E.,
  Watkins H.~J., 2008, \mnras, 393, 5

\bibitem[{{Bolton}(1975)}]{1975ApJ...200..269B}
{Bolton} C.~T., 1975, \apj, 200, 269

\bibitem[{{Bozzo} et~al.(2008{\natexlab{a}}){Bozzo}, {Falanga} \&
  {Stella}}]{2008ApJ...683.1031B}
{Bozzo} E., {Falanga} M., {Stella} L., 2008{\natexlab{a}}, \apj, 683, 1031

\bibitem[{{Bozzo} et~al.(2008{\natexlab{b}})}]{2008ATel.1493....1B}
{Bozzo} E., {Campana} S., {Stella} L., {Falanga} M., {Israel} G., {Rampy} R.,
  {Smith} D., {Negueruela} I., 2008{\natexlab{b}}, \atel, 1493, 1

\bibitem[{{Bozzo} et~al.(2012)}]{2012A&A...544A.118B}
{Bozzo} E., {Pavan} L., {Ferrigno} C., {Falanga} M., {Campana} S., {Paltani}
  S., {Stella} L., {Walter} R., 2012, \aap, 544, A118

\bibitem[{{Casares} et~al.(2014){Casares}, {Negueruela}, {Rib{\'o}}, {Ribas},
  {Paredes}, {Herrero} \& {Sim{\'o}n-D{\'{\i}}az}}]{2014Natur.505..378C}
{Casares} J., {Negueruela} I., {Rib{\'o}} M., {Ribas} I., {Paredes} J.~M.,
  {Herrero} A., {Sim{\'o}n-D{\'{\i}}az} S., 2014, \nat, 505, 378

\bibitem[{{Clark} et~al.(2010)}]{Clarke2010}
{Clark} D.~J. et~al., 2010, \mnras, 406, L75

\bibitem[{{Coleiro} \& {Chaty}(2013)}]{2013ApJ...764..185C}
{Coleiro} A., {Chaty} S., 2013, \apj, 764, 185

\bibitem[{{Corbet} et~al.(1999){Corbet}, {Finley} \&
  {Peele}}]{1999ApJ...511..876C}
{Corbet} R.~H.~D., {Finley} J.~P., {Peele} A.~G., 1999, \apj, 511, 876

\bibitem[{{Cowley} et~al.(1983){Cowley}, {Crampton}, {Hutchings}, {Remillard}
  \& {Penfold}}]{1983ApJ...272..118C}
{Cowley} A.~P., {Crampton} D., {Hutchings} J.~B., {Remillard} R., {Penfold}
  J.~E., 1983, \apj, 272, 118

\bibitem[{{Dickey} \& {Lockman}(1990)}]{1990ARA&A..28..215D}
{Dickey} J.~M., {Lockman} F.~J., 1990, \araa, 28, 215

\bibitem[{{Doroshenko} et~al.(2011){Doroshenko}, {Santangelo} \&
  {Suleimanov}}]{2011A&A...529A..52D}
{Doroshenko} V., {Santangelo} A., {Suleimanov} V., 2011, \aap, 529, A52

\bibitem[{{Drave} et~al.(2014){Drave}, {Bird}, {Sidoli}, {Sguera}, {Bazzano},
  {Hill} \& {Goossens}}]{2014MNRAS.439.2175D}
{Drave} S.~P., {Bird} A.~J., {Sidoli} L., {Sguera} V., {Bazzano} A., {Hill}
  A.~B., {Goossens} M.~E., 2014, \mnras, 439, 2175

\bibitem[{{Drave} et~al.(2013)}]{2013MNRAS.433..528D}
{Drave} S.~P., {Bird} A.~J., {Sidoli} L., {Sguera} V., {McBride} V.~A., {Hill}
  A.~B., {Bazzano} A., {Goossens} M.~E., 2013, \mnras, 433, 528

\bibitem[{{Enoto} et~al.(2014)}]{2014ApJ...786..127E}
{Enoto} T. et~al., 2014, \apj, 786, 127

\bibitem[{{Filliatre} \& {Chaty}(2004)}]{2004ApJ...616..469F}
{Filliatre} P., {Chaty} S., 2004, \apj, 616, 469

\bibitem[{{Gabriel} et~al.(2004)}]{2004ASPC..314..759G}
{Gabriel} C. et~al., 2004, in F.~{Ochsenbein}, M.G. {Allen}, D.~{Egret}, eds,
  \aspc. Astronomical Society of the Pacific Conference Series, Vol. 314, p.
  759

\bibitem[{{Goldwurm} et~al.(2003)}]{2003A&A...411L.223G}
{Goldwurm} A. et~al., 2003, \aap, 411, L223

\bibitem[{{Grebenev} \& {Sunyaev}(2005)}]{2005AstL...31..672G}
{Grebenev} S.~A., {Sunyaev} R.~A., 2005, \astlet, 31, 672

\bibitem[{{Grebenev} \& {Sunyaev}(2007)}]{2007AstL...33..149G}
{Grebenev} S.~A., {Sunyaev} R.~A., 2007, \astlet, 33, 149

\bibitem[{{Hill} et~al.(2005)}]{2005A&A...439..255H}
{Hill} A.~B. et~al., 2005, \aap, 439, 255

\bibitem[{{in 't Zand} et~al.(1998){in 't Zand}, {Heise}, {Smith}, {Muller},
  {Ubertini} \& {Bazzano}}]{1998IAUC.6840....2I}
{in 't Zand} J., {Heise} J., {Smith} M., {Muller} J.~M., {Ubertini} P.,
  {Bazzano} A., 1998, \iaucirc, 6840, 2

\bibitem[{{in't Zand}(2005)}]{2005A&A...441L...1I}
{in't Zand} J.~J.~M., 2005, \aap, 441, L1

\bibitem[{{Jansen} et~al.(2001)}]{2001A&A...365L...1J}
{Jansen} F. et~al., 2001, \aap, 365, L1

\bibitem[{{Kennea} et~al.(2014)}]{2014ATel.5980....1K}
{Kennea} J.~A. et~al., 2014, \atel, 5980, 1

\bibitem[{{Kreykenbohm} et~al.(2002){Kreykenbohm}, {Coburn}, {Wilms},
  {Kretschmar}, {Staubert}, {Heindl} \& {Rothschild}}]{2002A&A...395..129K}
{Kreykenbohm} I., {Coburn} W., {Wilms} J., {Kretschmar} P., {Staubert} R.,
  {Heindl} W.~A., {Rothschild} R.~E., 2002, \aap, 395, 129

\bibitem[{{Leahy} et~al.(1983){Leahy}, {Darbro}, {Elsner}, {Weisskopf}, {Kahn},
  {Sutherland} \& {Grindlay}}]{1983ApJ...266..160L}
{Leahy} D.~A., {Darbro} W., {Elsner} R.~F., {Weisskopf} M.~C., {Kahn} S.,
  {Sutherland} P.~G., {Grindlay} J.~E., 1983, \apj, 266, 160

\bibitem[{{Liu} et~al.(2006){Liu}, {van Paradijs} \& {van den
  Heuvel}}]{2006A&A...455.1165L}
{Liu} Q.~Z., {van Paradijs} J., {van den Heuvel} E.~P.~J., 2006, \aap, 455,
  1165

\bibitem[{{Lomb}(1976)}]{1976Ap&SS..39..447L}
{Lomb} N.~R., 1976, \apss, 39, 447

\bibitem[{{Manousakis} \& {Walter}(2015)}]{2015A&A...575A..58M}
{Manousakis} A., {Walter} R., 2015, \aap, 575, A58

\bibitem[{{Negueruela} \& {Smith}(2006)}]{2006ATel..831....1N}
{Negueruela} I., {Smith} D.~M., 2006, \atel, 831, 1

\bibitem[{{Negueruela} et~al.(2006){Negueruela}, {Smith}, {Reig}, {Chaty} \&
  {Torrej{\'o}n}}]{2006ESASP.604..165N}
{Negueruela} I., {Smith} D.~M., {Reig} P., {Chaty} S., {Torrej{\'o}n} J.~M.,
  2006, in A.~{Wilson}, ed., The X-ray Universe 2005. ESA Special Publication,
  Vol. 604, p. 165

\bibitem[{{Negueruela} et~al.(2008){Negueruela}, {Torrej{\'o}n}, {Reig},
  {Rib{\'o}} \& {Smith}}]{2008AIPC.1010..252N}
{Negueruela} I., {Torrej{\'o}n} J.~M., {Reig} P., {Rib{\'o}} M., {Smith} D.~M.,
  2008, in R.M. {Bandyopadhyay}, S.~{Wachter}, D.~{Gelino}, C.R. {Gelino}, eds,
  A Population Explosion: The Nature \& Evolution of X-ray Binaries in Diverse
  Environments. American Institute of Physics Conference Series, Vol. 1010, pp.
  252--256

\bibitem[{{Odaka} et~al.(2013){Odaka}, {Khangulyan}, {Tanaka}, {Watanabe},
  {Takahashi} \& {Makishima}}]{2013ApJ...767...70O}
{Odaka} H., {Khangulyan} D., {Tanaka} Y.~T., {Watanabe} S., {Takahashi} T.,
  {Makishima} K., 2013, \apj, 767, 70

\bibitem[{{Paizis} \& {Sidoli}(2014)}]{Paizis2014}
{Paizis} A., {Sidoli} L., 2014, \mnras, 439, 3439

\bibitem[{{Press} \& {Rybicki}(1989)}]{1989ApJ...338..277P}
{Press} W.~H., {Rybicki} G.~B., 1989, \apj, 338, 277

\bibitem[{{Reynolds} et~al.(1999){Reynolds}, {Owens}, {Kaper}, {Parmar} \&
  {Segreto}}]{1999A&A...349..873R}
{Reynolds} A.~P., {Owens} A., {Kaper} L., {Parmar} A.~N., {Segreto} A., 1999,
  \aap, 349, 873

\bibitem[{{Romano} et~al.(2011)}]{2011MNRAS.412L..30R}
{Romano} P. et~al., 2011, \mnras, 412, L30

\bibitem[{{Romano} et~al.(2014)}]{2014A&A...562A...2R}
{Romano} P. et~al., 2014, \aap, 562, A2

\bibitem[{{Romano} et~al.(2015)}]{2015A&A...576L...4R}
{Romano} P. et~al., 2015, \aap, 576, L4

\bibitem[{{Scargle}(1982)}]{1982ApJ...263..835S}
{Scargle} J.~D., 1982, \apj, 263, 835

\bibitem[{{Sguera} et~al.(2005)}]{Sguera2005}
{Sguera} V. et~al., 2005, \aap, 444, 221

\bibitem[{{Sguera} et~al.(2006)}]{2006ApJ...646..452S}
{Sguera} V. et~al., 2006, \apj, 646, 452

\bibitem[{Shakura et~al.(2012)Shakura, Postnov, Kochetkova \&
  Hjalmarsdotter}]{Shakura2012a}
Shakura N., Postnov K., Kochetkova A., Hjalmarsdotter L., 2012, \mnras, 420,
  216

\bibitem[{{Shakura} et~al.(2013){Shakura}, {Postnov} \&
  {Hjalmarsdotter}}]{2013MNRAS.428..670S}
{Shakura} N., {Postnov} K., {Hjalmarsdotter} L., 2013, \mnras, 428, 670

\bibitem[{{Shakura} et~al.(2014){Shakura}, {Postnov}, {Sidoli} \&
  {Paizis}}]{2014MNRAS.442.2325S}
{Shakura} N., {Postnov} K., {Sidoli} L., {Paizis} A., 2014, \mnras, 442, 2325

\bibitem[{{Sidoli} et~al.(2007){Sidoli}, {Romano}, {Mereghetti}, {Paizis},
  {Vercellone}, {Mangano} \& {G{\"o}tz}}]{2007A&A...476.1307S}
{Sidoli} L., {Romano} P., {Mereghetti} S., {Paizis} A., {Vercellone} S.,
  {Mangano} V., {G{\"o}tz} D., 2007, \aap, 476, 1307

\bibitem[{{Sidoli} et~al.(2009)}]{2009MNRAS.400..258S}
{Sidoli} L. et~al., 2009, \mnras, 400, 258

\bibitem[{{Sidoli} et~al.(2012){Sidoli}, {Mereghetti}, {Sguera} \&
  {Pizzolato}}]{2012MNRAS.420..554S}
{Sidoli} L., {Mereghetti} S., {Sguera} V., {Pizzolato} F., 2012, \mnras, 420,
  554

\bibitem[{{Sidoli} et~al.(2013)}]{2013MNRAS.429.2763S}
{Sidoli} L. et~al., 2013, \mnras, 429, 2763

\bibitem[{{Smith} et~al.(2012){Smith}, {Markwardt}, {Swank} \&
  {Negueruela}}]{2012MNRAS.422.2661S}
{Smith} D.~M., {Markwardt} C.~B., {Swank} J.~H., {Negueruela} I., 2012, \mnras,
  422, 2661

\bibitem[{{Str{\"u}der} et~al.(2001)}]{2001A&A...365L..18S}
{Str{\"u}der} L. et~al., 2001, \aap, 365, L18

\bibitem[{{Torrej{\'o}n} et~al.(2010){Torrej{\'o}n}, {Negueruela}, {Smith} \&
  {Harrison}}]{2010A&A...510A..61T}
{Torrej{\'o}n} J.~M., {Negueruela} I., {Smith} D.~M., {Harrison} T.~E., 2010,
  \aap, 510, A61

\bibitem[{{Turner} et~al.(2001)}]{2001A&A...365L..27T}
{Turner} M.~J.~L. et~al., 2001, \aap, 365, L27

\bibitem[{{Ubertini} et~al.(2003)}]{2003A&A...411L.131U}
{Ubertini} P. et~al., 2003, \aap, 411, L131

\bibitem[{{Walter} et~al.(2003)}]{2003A&A...411L.427W}
{Walter} R. et~al., 2003, \aap, 411, L427

\bibitem[{{Walter} et~al.(2006)}]{2006A&A...453..133W}
{Walter} R. et~al., 2006, \aap, 453, 133

\bibitem[{{Wilms} et~al.(2000){Wilms}, {Allen} \&
  {McCray}}]{2000ApJ...542..914W}
{Wilms} J., {Allen} A., {McCray} R., 2000, \apj, 542, 914

\bibitem[{{Winkler} et~al.(2003)}]{2003A&A...411L...1W}
{Winkler} C. et~al., 2003, \aap, 411, L1

\bibitem[{Zurita-Heras \& Chaty(2008)}]{Heras2008}
Zurita-Heras J.~A., Chaty S., 2008, \aap, 493, 4

\end{thebibliography}

\label{lastpage}
\end{document}